\documentclass[fleqn,usenatbib]{mnras}

\usepackage{newtxtext,newtxmath}
 
\usepackage[T1]{fontenc}

\DeclareRobustCommand{\VAN}[3]{#2}
\let\VANthebibliography\thebibliography
\def\thebibliography{\DeclareRobustCommand{\VAN}[3]{##3}\VANthebibliography}

\usepackage{graphicx}	% Including figure files
\usepackage{amsmath}	% Advanced maths commands
\usepackage{amssymb}	% Extra maths symbols
\usepackage{subfigure}
\newcommand{\bs}{\boldsymbol}
\newcommand{\percent}{\,\mathrm{per\,cent}}
\everymath{\displaystyle}

\newcommand*{\GtrSim}{\smallrel\gtrsim}

\makeatletter
\newcommand*{\smallrel}[2][.8]{%
  \mathrel{\mathpalette{\smallrel@{#1}}{#2}}%
}
\newcommand*{\smallrel@}[3]{%
  \sbox0{$#2\vcenter{}$}%
  \dimen@=\ht0 %
  \raise\dimen@\hbox{%
    \scalebox{#1}{%
      \raise-\dimen@\hbox{$#2#3\m@th$}%
    }%
  }%
}
\makeatother

\usepackage{pdflscape}
\usepackage{hhline}
\usepackage{setspace}
\usepackage[flushleft]{threeparttable}
\title[Near-Gaussian distributions for discrete data]{Near-Gaussian distributions for modelling discrete stellar velocity data with heteroskedastic uncertainties}

\author[Jason L. Sanders \& N. Wyn Evans]{
Jason L. Sanders,$^{1,2}$\thanks{JLS (jason.sanders@ucl.ac.uk), NWE (nwe@ast.cam.ac.uk)}
and N. Wyn Evans$^{1}$
\\
$^{1}$Institute of Astronomy, University of Cambridge, Madingley Rise, Cambridge, CB3 0HA, UK\\
$^{2}$Department of Physics and Astronomy, University College London, London WC1E 6BT, UK
}

\date{Accepted XXX. Received YYY; in original form ZZZ}

\pubyear{2020}

\begin{document}
\label{firstpage}
\pagerange{\pageref{firstpage}--\pageref{lastpage}}
\maketitle

% Abstract of the paper
\begin{abstract}
The velocity distributions of stellar tracers in general exhibit weak non-Gaussianity encoding information on the orbital composition of a galaxy and the underlying potential. The standard solution for measuring non-Gaussianity involves constructing a series expansion (e.g. the Gauss-Hermite series) which can produce regions of negative probability density. This is a significant issue for the modelling of discrete data with heteroskedastic uncertainties. Here, we introduce a method to construct positive-definite probability distributions by the convolution of a given kernel with a Gaussian distribution. Further convolutions by observational uncertainties are trivial.
The statistics (moments and cumulants) of the resulting distributions are governed by the kernel distribution. Two kernels (uniform and Laplace) offer simple drop-in replacements for a Gauss-Hermite series for negative and positive excess kurtosis distributions with the option of skewness. We demonstrate the power of our method by an application to real and mock line-of-sight velocity datasets on dwarf spheroidal galaxies, where kurtosis is indicative of orbital anisotropy and hence a route to breaking the mass-anisotropy degeneracy for the identification of cusped versus cored dark matter profiles. Data on the Fornax dwarf spheroidal galaxy indicate positive excess kurtosis and hence favour a cored dark matter profile. Although designed for discrete data, the analytic Fourier transforms of the new models also make them appropriate for spectral fitting, which could improve the fits of high quality data by avoiding unphysical negative wings in the line-of-sight velocity distribution.
\end{abstract}

% Select between one and six entries from the list of approved keywords.
% Don't make up new ones.

%%%%%%%%%%%%%%%%%%%%%%%%%%%%%%%%%%%%%%%%%%%%%%%%%%

%%%%%%%%%%%%%%%%% BODY OF PAPER %%%%%%%%%%%%%%%%%%
\section{Introduction}
The velocity distributions of stars within a galaxy exhibit a range of morphologies. Non-Gaussianity, particularly kurtosis, in the velocity distributions is linked to anisotropy of the orbits of the stars: populations of stars on more radial orbits typically produce `peakier' or long-tailed velocity distributions, whilst populations of stars on circular orbits produce more flat-topped, short-tailed distributions \citep{Gerhard1993,vanderMarelFranx1993}. Furthermore, rotating populations generically produce skewed distributions, as demonstrated by the azimuthal velocities of the Milky Way disc populations~\citep[e.g., Chap. 10,][]{BM} or the $h_3$-$V/\sigma$ relation observed in elliptical galaxies \citep{Bender1994}. However, in the main, these deviations from Gaussianity are weak and so can be captured via a Gauss-Hermite series where the zeroth-order model is a Gaussian and orthogonal polynomial corrections produce non-Gaussianity. Typically, only two correction terms are used with coefficients $h_3$ quantifying the skewness and $h_4$ quantifying the (excess) kurtosis \citep{vanderMarelFranx1993,Gerhard1993}.

Gauss-Hermite coefficients have seen wide-spread use in the summary of integral field unit spectroscopy data \citep{SAURON,ATLAS3D,SAMI} where typically one simultaneously extracts information on the stellar populations and their kinematics via template fitting \citep{CappellariEmsellem2004,Cappellari2017}. The coefficients provide useful intermediate data products for comparison with dynamical models \citep[e.g.][]{Rix1997}, whilst also giving physical insight into the orbital distributions and hence potential. They were originally developed to characterise the velocity profiles near the centres of elliptical galaxies and hence to measure the masses of central black holes~\citep[e.g.][]{vdM94,vanderMarel1998}. Applications to barred galaxies have also been suggested. \cite{Debattista2005} identified negative $h_4$ (negative excess kurtosis) as an indicator of viewing a face-on peanut-shaped bar/bulge \citep[e.g.][]{MendezAbreu2019}. Recently, \cite{SellwoodGerhard2020} have suggested the sign of $h_4$ of the vertical velocity distribution distinguishes different formation channels for a peanut-shaped bar/bulge with positive $h_4$ (longer tails) indicative of a peanut formed through resonances, whilst negative $h_4$ (flat-topped distribution) indicative of a more violent buckling event which depletes the bar of low vertical velocity stars.

Higher-order moments have also been used in the analysis of dwarf spheroidal galaxy line-of-sight velocity data. 
Typically, they have been used to address the question of the nature of the dark matter density at the centre -- whether it is cusped like the famous \citet*{NFW} model or cored like a harmonic potential. This information is summarised in the excess kurtosis \citep{Lokas2002} and can be extracted from data using Gauss-Hermite coefficients \citep{AmoriscoEvans2012}, the virial shape parameters \citep{MerrifieldKent1990,Read2019}, higher order Jeans equations \citep{Richardson2013} or through the use of a dynamical distribution functions \citep{Lokas2002,Lokas2005,AE11, Breddels2013,Pascale2018}. 

In the application to discrete velocity measurements with heteroskedastic uncertainties, as for dwarf spheroidals, we hit an immediate stumbling block when using the Gauss-Hermite series: in general a Gauss-Hermite series is not positive definite and so cannot be interpreted as a probability density function. This means it cannot be used in a probabilistic framework which incorporates observational uncertainties or membership probabilities (in a mixture model). Although unphysical, typically the regions of negativity are small, so for spectral fitting this isn't a significant issue. However, negative probabilities in any probabilistic framework produce awkward numerical problems.
For the case of dwarf spheroidals, there are datasets of at most a few thousand stars \citep[e.g.][]{Walker2009} and the varying stellar types produce a range of differing uncertainties with which velocities can be measured. In order to extract the maximal information from such a dataset, we wish to avoid either binning the data or assuming the uncertainties are small. 
\cite{KuijkenMerrifield1993} explored the possibility of ensuring positive-definiteness by using a Gaussian mixture, which also allows for simple convolution with measurement uncertainties.
\cite{AmoriscoEvans2012} provided an alternative solution to the issue by constructing probability density functions inspired by dynamical galaxy models. This approach produced expressions that required numerical integration, although the authors demonstrate how to make their procedure numerically efficient. Here, we provide an alternative and simpler method for the construction of appropriate analytic probability density functions. Although slightly more restrictive, they offer a drop-in replacement for the Gauss-Hermite series which guarantees a positive-definite probability density function. Here, we focus on data from dwarf spheroidals, but our methods can be applied to any sets of discrete tracers such as globular clusters and planetary nebulae \citep{Ag14,Oldham2016} or clusters of galaxies \citep{Wojtak2010}.

We begin in Section~\ref{sec::GHSERIES} by describing how the Gauss-Hermite series can be used to model discrete tracers with heteroskedastic uncertainties and highlight its limitations, before presenting an alternative method for constructing appropriate non-Gaussian pdfs in Section~\ref{section::general_family}. We provide specific examples of the new families in Section~\ref{section::kernels} and apply the new probability density functions to the problem of dwarf spheroidal galaxies, specifically Fornax, in Section~\ref{section::application}. We close with a brief discussion of the use of the new models in the context of spectral fitting in Section~\ref{section::spectral} where the guaranteed non-negativity could potentially improve fits to high-quality data.

\section{The Gauss-Hermite series}\label{sec::GHSERIES}

An approximation to non-Gaussian velocity distributions is given by the Gauss-Hermite series \citep{Gerhard1993,vanderMarelFranx1993}
\begin{equation}
    f(x) = \frac{\lambda\alpha(w)}{\sigma}\Big(1+\sum_{i\geq3}h_i H_i(w)\Big);\quad w = \frac{x-V}{\sigma}.
    \label{eqn::gauss_hermite}
\end{equation}
Here, $H_i(w)$ are Hermite polynomials and 
\begin{equation}
    \alpha(w)=\frac{1}{\sqrt{2\pi}}e^{-\tfrac{1}{2}w^2},
\label{eqn::unitGaussian}
\end{equation}
is the weight function. The Hermite polynomials satisfy the orthogonality condition
\begin{equation}
    \int_{-\infty}^\infty\mathrm{d}w\,H_m(w)H_n(w)\alpha^2(w) = \frac{1}{\sqrt{4\pi}}\delta_{mn}.
\end{equation}
In Appendix~\ref{appendix::hermite}, we give the expression for the $n$th Hermite polynomial as well as the explicit formulae for the first five. The Gauss-Hermite series is similar to the Gram-Charlier and Edgeworth series familiar from the theory of statistics~\citep{KS} and discussed in Appendix~\ref{appendix::other_series}.

Typically, only two terms in the series~\eqref{eqn::gauss_hermite} are used giving five fitting parameters $p=(\lambda, V, \sigma, h_3, h_4)$. The first three terms give the amplitude, mean and dispersion of the distribution, whilst $h_3$ describes the skewness and $h_4$ the excess kurtosis (positive $h_4$ produces broader tails, whilst negative $h_4$ produces more truncated tails). Although $V$ and $\sigma$ are not strictly equal to the mean and dispersion for non-zero $h_i$, by the orthogonality properties of the series, they are the mean and dispersion of the best Gaussian fit. As the series is constructed from orthogonal polynomials, for some parameter choices the Gauss-Hermite series can have regions of negativity \citep[for example, see fig. 1 of ][]{vanderMarelFranx1993}. Assuming we can interpret $f(x)$ as a pdf, the choice of $\lambda$ which normalizes $f(x)$ such that $\int\mathrm{d}x\,f(x)=1$ is $\lambda=(1+\sqrt{3/8}h_4)^{-1}$. 

\subsection{Fitting a Gauss-Hermite series to data}

The Gauss-Hermite series was introduced as an approximation to a galaxy's line-of-sight velocity distribution (LOSVD) as measured from spectra \citep{Gerhard1993,vanderMarelFranx1993}. Given a profile $\mathcal{L}(x)$ measured with uncertainty $\sigma_\mathcal{L}(x)$, the best-fitting parameters $p$
are measured by minimising
\begin{equation}
\int\mathrm{d}x\,\frac{(\mathcal{L}(x)-f(x))^2}{\sigma^2_\mathcal{L}(x)}.
\label{eqn::chi_cont}
\end{equation}
In practice, this integral is performed as a sum over spectral pixels. For a model galaxy template spectrum, the kinematics are incorporated through convolution with a Gauss-Hermite series \citep{CappellariEmsellem2004}. 
As discussed by \cite{Cappellari2017}, when the dispersion is smaller than half the pixel spacing, this convolution is most efficiently performed by first taking the analytic Fourier transform of the Gauss-Hermite series, before taking the discrete inverse transform of the product of the Fourier template and the Fourier Gauss-Hermite series. Further Gaussian broadening is trivially incorporated in Fourier space. 
In these methods, the requirement of a well-defined probability density function ($f(x)>0$ everywhere) is unimportant as it does not significantly alter the computation of equation~\eqref{eqn::chi_cont}. In these cases, it is possible to introduce a truncation by setting $f(x)=0$ when $f(x)<0$ which can improve the fit for $f(x)>0$ \citep{vanderMarelFranx1993}.

When we have discrete velocity measurements $\{x_i\}$, as is common for Milky Way or Local Group studies, these fitting methods are not always easily applicable. One simple method applicable for error-free data is to relate $h_3$ and $h_4$ to moments of the data. As shown in Appendix~\ref{appendix::gh_cumulants}, for small deviations from Gaussianity, $h_3$ and $h_4$ are related to the skewness $g$ and excess kurtosis $\kappa$ as 
\begin{equation}
    h_3 = \frac{g}{4\sqrt{3}}, \:
    h_4 = \frac{\kappa}{8\sqrt{6}}.
    \label{eqn::h3h4_skewkurt}
\end{equation}
For $N$ samples, the corresponding variance in these quantities (for near Gaussian distributions) are
\citep{Kenney}
\begin{equation}
    \begin{split}
        \mathrm{Var}(
         h_3)&=\frac{N(N-1)}{8(N-2)(N+1)(N+3)},\\
        \mathrm{Var}( h_4)&=\frac{N(N-1)^2}{16(N-3)(N-2)(N+3)(N+5)}.
    \end{split}
\label{eqn::standard_errors}
\end{equation}
Typically, however, equation~\eqref{eqn::h3h4_skewkurt} gives quite poor approximations for $h_3$ and $h_4$ (in particular $h_4$ where it is only valid to $10\percent$ for $|h_4|<0.01$). In turn, equation~\eqref{eqn::standard_errors} gives poor estimates of the excepted error and only appears valid using Gaussian-distributed samples of size $\GtrSim1000$. For smaller samples, following \cite{AmoriscoEvans2012}, the uncertainty in $h_3$ and $h_4$ is better approximated as $(2N)^{-1/2}$.

A further method is to bin the data into an approximation of $\mathcal{L}(x)$ and minimise equation~\eqref{eqn::chi_cont}. This is undesirable, as it generically loses information through smoothing and typically we only have a small numbers of measurements to work with. 
\cite{SellwoodGerhard2020} propose a method for avoiding binning where the (analytic) cumulative distribution function for the Gauss-Hermite series is fitted to the cumulative velocity distribution of simulation data. Such a method does not adapt well when observational uncertainties
are considered. 

To avoid binning and fully incorporate the measurement uncertainties for a set of samples $\{x_i\}$ measured with uncertainties $\{\sigma_{ei}\}$, we should maximise
%
% \begin{equation}
%     \prod_i\int_{-\infty}^{\infty}\mathrm{d}x'\,f(x')
%     (2\pi \sigma_{ei}^2)^{-1/2}e^{-(x_i-x')^2/(2\sigma_{ei}^2)}.
% \end{equation}
% %
\begin{equation}
    \prod_i\int_{-\infty}^{\infty}\mathrm{d}x'\,f(x')\mathcal{N}(x_i-x'|\sigma_{ei}),
\end{equation}
where
\begin{equation}
    \mathcal{N}(x|\sigma)
    =(2\pi\sigma^2)^{-1/2}e^{-\tfrac{1}{2}x^2/\sigma^2}.
\end{equation}
Generically, the Gauss-Hermite series (and other series expansions) produces negative values of $f(x)$ for some choice of $x$.
% , particularly for large values of $|h_3|$ and $|h_4|$. 
This makes the interpretation of $f(x)$ as a probability distribution function (pdf) awkward. Regions of negative probability are particularly an issue for negative $h_4$, where the pdf is zero somewhere for any reasonable choice of $h_3$. However, usually the region of negative pdf is small. This deficiency is not simple to fix with the addition of higher order terms. An alternative is to truncate the pdf at the first zero crossing. The roots of the Gauss-Hermite series for $h_3=0$ are 
\begin{equation}
w=\frac{3}{2}\pm\frac{1}{\sqrt{2}h_4}\sqrt{3h_4^2-\sqrt{6}h_4}.
\end{equation}
It is then possible to analytically normalize $f(x)$ between these limits, although the expressions are long-winded. Incorporating these workarounds into a likelihood call significantly increases the cost of a likelihood evaluation and is not easily adapted for $h_3\neq0$.

In the next subsection, we will demonstrate how the Gauss-Hermite series can be used when $f(x)$ is a well-defined probability density function, before we move on to presenting an alternative set of models. 

\subsection{Convolution of the Gauss-Hermite series with observational uncertainties}

A sample of points $x_i$ drawn from $f(x)$ but measured with some uncertainty $\sigma_e$ follow the distribution $f_{\sigma_e}(x)$ given by
\begin{equation}
    f_{\sigma_e}(x)=\int_{-\infty}^{\infty}\mathrm{d}x'\,f(x')
    \mathcal{N}(x-x'|\sigma_e).
    \label{eqn::convGH}
\end{equation}
Given the form of the Gauss-Hermite series, it is possible to compute this convolution analytically. In Appendix~\ref{appendix::convolution} we provide the details and here quote the final result as
\begin{equation}
\begin{split}
    f_{\sigma_e}&(x) = \frac{\lambda}{\sigma'}\alpha(w')\Big\{1+\\&\sum_{n\geq3}h_n\Big(\frac{\sigma}{\sigma'}\Big)^{n}\sum _ { j = 0 } ^ { \lfloor \frac { n } { 2 } \rfloor }\Big(\frac{\sigma_e}{\sigma}\Big)^{2j} 
    \frac{\sqrt{n!}}{j!2^{j}\sqrt{(n-2j)!}}
    H _ { n - 2 j }(w')\Big\},
\end{split}
\label{eqn::convGH_fullresult}
\end{equation}
where $w'=(x-V)/\sigma'$ and $\sigma'=\sqrt{\sigma^2+\sigma_e^2}$. For the truncated series $h_i=0$ for $i>4$, we find 
\begin{equation}
\begin{split}
    f_{\sigma_e}(x&)= \frac{\lambda}{\sigma'}\alpha(w')\Big\{1+\\&\frac{h_3}{(\sigma'/\sigma)^3}\Big[H_3(w')+\sqrt{\frac{3}{2}}\Big(\frac{\sigma_e}{\sigma}\Big)^2H_1(w')\Big]+\\&\frac{h_4}{(\sigma'/\sigma)^4}\Big[H_4(w')+\sqrt{3}\Big(\frac{\sigma_e}{\sigma}\Big)^2H_2(w')+\sqrt{\frac{3}{8}}\Big(\frac{\sigma_e}{\sigma}\Big)^4H_0(w')\Big]\Big\}.
    \label{eqn::convGH_result}
\end{split}
\end{equation}
This function can now be used as the likelihood of data with heteroskedastic errors $\sigma_{ei}$ to recover the underlying $h_3$ and $h_4$ moments, provided $f(x)$ is everywhere positive. 

% \subsubsection{Gauss-Hermite coefficient biases}

\begin{figure}
    \centering
    \includegraphics[width=\columnwidth]{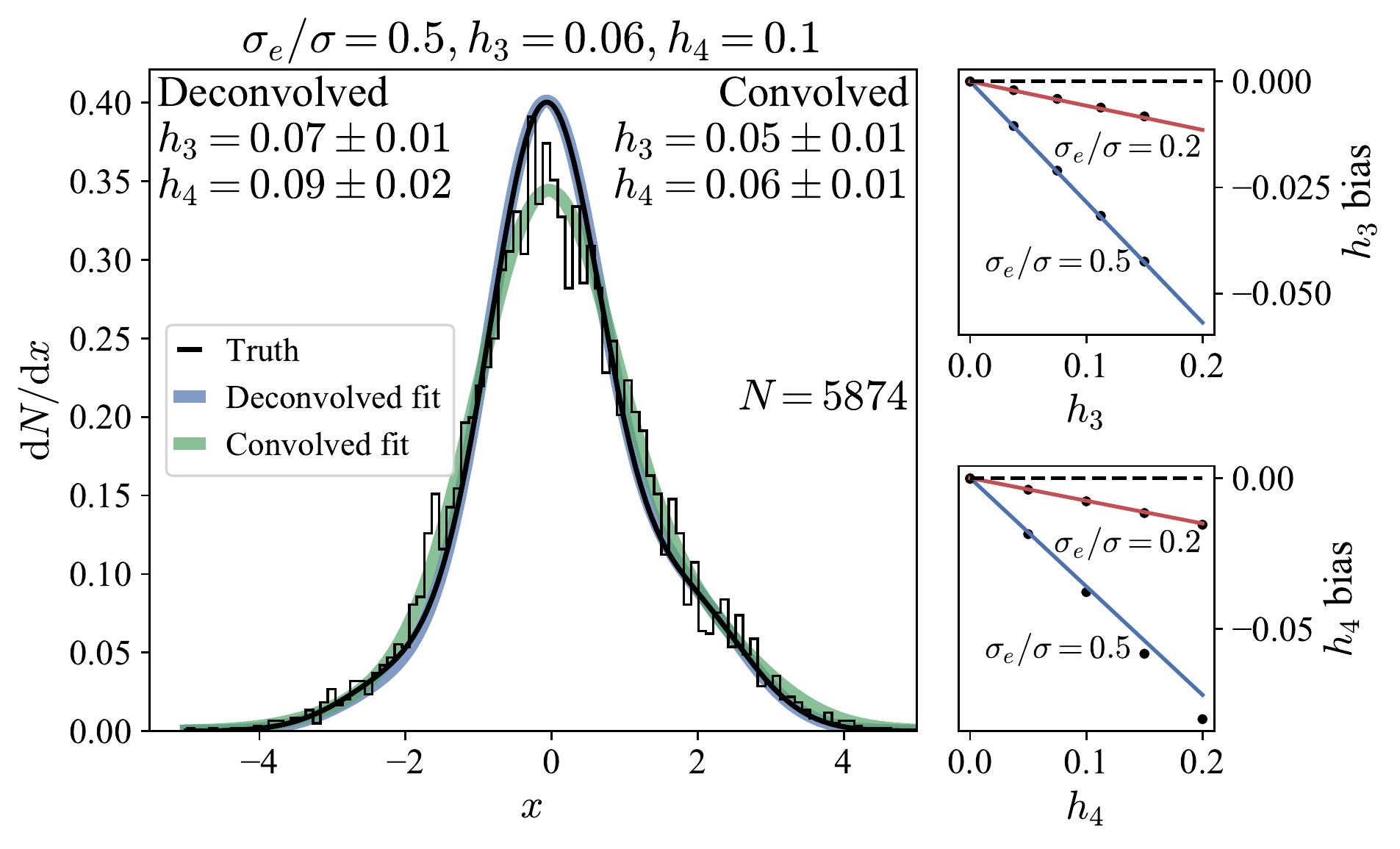}
    \caption{Gauss-Hermite series fitting to a mock distribution: the left panel shows a sample of $N$ draws (the black histogram) from a Gauss-Hermite series with $h_3=0.06$ and $h_4=0.1$ (the black line). The samples have then been scattered by an uncertainty of $0.5$. The blue line is the fit result accounting for uncertainties whilst green is without accounting for uncertainties. Note that the recovered $h_3$ and $h_4$ are higher when the uncertainties are accounted for. This is also depicted in the right panels which show the bias in $h_3$ and $h_4$ as a function of these parameters if one fails to account for the uncertainties. Two choices of error are shown ($0.2$ and $0.5$). The lines give the approximations $h_3((1+(\sigma_e/\sigma)^2)^{-3/2}-1)$ and $h_4((1+(\sigma_e/\sigma)^2)^{-2}-1)$.}
    \label{fig:h3h4_bias}
\end{figure}

Equation~\eqref{eqn::convGH_result} can be used to assess the impact of uncertainties on the recovery of the Gauss-Hermite coefficients.
We observe for $h_3=h_4=0$ we recover the standard result that the convolution broadens the Gaussian from dispersion $\sigma$ to $\sigma'$. Note that for non-zero $h_3$ and $h_4$, the convolution also introduces contributions from $H_1(w')$ and $H_2(w')$ which we identify as further modifications to the mean and dispersion. However, in the limit of small uncertainties $\sigma_e\ll\sigma$, we find
\begin{equation}
\begin{split}
    f_{\sigma_e}(x) &\approx \frac{\lambda}{\sigma'}\alpha(w')\Big(1+h_3'H_3(w')+h_4'H_4(w')\Big),
\end{split}
\end{equation}
where $h_3'=h_3(\sigma/\sigma')^3$ and $h_4'=h_4(\sigma/\sigma')^4$. Therefore, small uncertainties don't bias the mean, but bias the dispersion in the expected way: $\sigma'^2=\sigma^2+\sigma_e^2$. We remark that $h_3$ and $h_4$ are also biased: without accounting for uncertainties $h_3$ is reduced by a factor $(1+(\sigma_e/\sigma)^2)^{-3/2}$ and $h_4$ by a factor of $(1+(\sigma_e/\sigma)^2)^{-2}$. 
In Fig.~\ref{fig:h3h4_bias}, we show the result of fitting a Gauss-Hermite series with $h_3=0.06$ and $h_4=0.1$ to a sample of $N\approx6000$ datapoints scattered by two choices of uncertainty $\sigma_e/\sigma=0.2$ and  $\sigma_e/\sigma=0.5$. We see the true parameters are recovered when sampling from the likelihood $\textstyle \sum_i\ln f_{\sigma_{ei}}(x_i)$ using the \textsc{emcee} sampler \citep{emcee}. If one fails to account for the uncertainties, we recover the expected biases in $h_3$ and $h_4$ \citep[see][for a similar discussion]{AmoriscoEvans2012}.

We have demonstrated how the Gauss-Hermite series can be used to model discrete data with heteroskedastic uncertainties, but only when the series is positive definite. Generically, for negative $h_4$ the series will produce negative wings making its use in a probabilistic framework awkward. In the next section, we will introduce a method to produce everywhere positive-definite pdfs which can be used as replacements for the Gauss-Hermite series.

\section{A general family of weakly non-Gaussian pdfs}\label{section::general_family}

Our goal is to construct a family of everywhere positive pdfs, which can model weakly non-Gaussian distributions and also are able to incorporate observational uncertainties. This means that the models must have an easily computable convolution with a Gaussian kernel. This class of models is very desirable for modelling star-by-star data with heteroskedastic observational uncertainties, as otherwise the likelihood call for each star requires an integration over the uncertainty. We begin with a general scheme for constructing such models before discussing in detail two specific examples useful for modelling positive and negative excess kurtosis. 

\subsection{Model specification}

With the requirement of analytic convolution with uncertainties, it is logical to investigate models which have been constructed by a convolution of a general kernel with a Gaussian. Further Gaussian kernel convolutions can then be simply performed and transform the models into another Gaussian convolution with a slightly modified general kernel. The models are
\begin{equation}
    f(w)=b\int_{-\infty}^\infty \mathrm{d}y\,K(y)\alpha(y-b(w-w_0)),
\end{equation}
where $\alpha(y)$ is a unit Gaussian as defined in equation~\eqref{eqn::unitGaussian} and $K(y)$ is a general kernel normalized 
to have unit weight. $w_0$ and $b$ are parameters controlling the mean and variance of the model. To reproduce the unit Gaussian model $\alpha(w)$ when $K(y)=\delta(y)$, we require $b=1$ and $w_0=0$ in this limit, which implies $b$ and $w_0$ in general depend upon the parameters of the kernel $K(y)$. In the opposite limit where the kernel $K(y)$ is significantly broader than the unit Gaussian, $\alpha(y-b(w-w_0))\rightarrow\delta(y-b(w-w_0))$ so $f(w)\rightarrow bK(b(w-w_0))$, a scaled and shifted version of the kernel. 

As with the Gauss-Hermite series, we have the freedom to introduce an arbitrary shift and scale to redefine the pdf in terms of $x=V+\sigma w$.

\subsubsection{Model moments and cumulants}

The moments of these models are given by
\begin{equation}
\begin{split}
    \mu_n(c) 
    &= \int_{-\infty}^\infty\mathrm{d}w\,(w-c)^n f(w),
    \\&= \frac{1}{b^n}\sum_{k=0}^{\lfloor \frac{n}{2}\rfloor}\binom{n}{2k}(2k-1)!!\tilde\mu_{n-2k}(bc-bw_0),
\end{split}
\end{equation}
where $\tilde\mu_n(c)$ are the moments of $K(y)$ (we use a tilde to denote properties of $K(y)$).
For instance, the mean $\mu\equiv\mu_1(0)$ is given by
\begin{equation}
\mu = \frac{1}{b}\int_{-\infty}^\infty\mathrm{d}y\,yK(y-bw_0)= \frac{\tilde\mu}{b}+w_0,
\end{equation}
where $\tilde\mu\equiv\tilde\mu(0)$.
The central moments are defined as $\mu_n\equiv\mu_n(\mu)$, which can be combined in the usual way to find the cumulants. However, the cumulants $\kappa_n$ are more simply computed via the cumulant generating function $\ln\phi(u)$ defined as
\begin{equation}
    \ln\phi(u)=\ln\int_{-\infty}^\infty\mathrm{d}w\,e^{iuw}f(w) = \sum_{r=1}^{r=\infty}\kappa_r\frac{(iu)^r}{r!}.
\end{equation}
In this way, the $r$th cumulant is
\begin{equation}
    \kappa_r = \frac{1}{i^r}\frac{\mathrm{d}^r\ln\phi(u)}{\mathrm{d}u^r}\Big|_{u=0}.
\end{equation}
From the properties of the Fourier transform, the cumulant generating function of $f(w)$ is given by
\begin{equation}
    \ln\phi(u) = iw_0u-\frac{u^2}{2b^2} + \ln \phi_K(u/b),
\end{equation}
where $\phi_K(u)$ is the characteristic function of $K(y)$,
\begin{equation}
    \phi_K(u)=\int_{-\infty}^\infty\mathrm{d}w\,e^{iuy}K(y).
\end{equation}
From the cumulants, we define the variance $v\equiv\kappa_2$, the skewness $g\equiv\kappa_3/v^{3/2}$ and the excess kurtosis $\kappa\equiv\kappa_4/v^2$.
We note that the variance of $K(y)$, $\tilde v$, is related to the variance of $f(w)$ by
\begin{equation}
    v = \frac{1}{b^2}(\tilde v+1),
\end{equation}
and all higher order cumulants ($r\geq3$) satisfy $\kappa_r=\tilde\kappa_r/b^r$.
This demonstrates that the signs of the skewness and the excess kurtosis of the kernel govern respectively the signs of the skewness and the excess kurtosis of the model.

\subsubsection{Convolution with uncertainties}

With such models, the convolution with observational uncertainties of magnitude $s$ is
\begin{equation}
\begin{split}
    f_s(w)&=\int_{-\infty}^\infty\mathrm{d}w'f(w')\mathcal{N}(w-w'|s) \\&= b\int_{-\infty}^\infty\mathrm{d}y\,K(ty)\alpha(y-b(w-w_0)/t),
\end{split}
\label{eqn::general_conv_errors}
\end{equation}
where $t^2=1+b^2s^2$ and $f(w)$ is recovered for $t=1$ (i.e. $f_0(w)=f(w)$). In principle, if the convolution with $K(y)$ is analytic, the convolution with $K(ty)$ is likely to be analytic. As the convolution is a multiplication in Fourier space, the cumulant generating function of the error-convolved model is given by
\begin{equation}
    \ln\phi_s(u)=\ln\phi(u)-\frac{u^2s^2}{2}.
\end{equation}

\subsubsection{Gauss-Hermite coefficients}
The (unnormalized) Gauss-Hermite coefficients of this class of models are given by
\begin{equation}
    h_n 
    = \sqrt{4\pi}\int_{-\infty}^\infty\mathrm{d}w\,\alpha(w)H_n(w)f(w).
\label{eqn::GH_mom}
\end{equation}
Note that in the limit where the kernel $K$ is significantly broader than the unit Gaussian,
the Gauss-Hermite coefficients are given by
\begin{equation}
h_n \rightarrow \sqrt{4\pi}b\int_{-\infty}^\infty\mathrm{d}y\,K(b(y-w_0))\alpha(y)H_n(y),
\label{eqn::wide_hn}
\end{equation}
i.e. they are the Gauss-Hermite coefficients of $K(b(y-w_0))$. 
A general expression for the Gauss-Hermite coefficients is found via the generating function of the Hermite polynomials given in equation~\eqref{eqn::hp_gf}. We compute the integral
\begin{equation}
\begin{split}
    \sum_{n=0}^{n=\infty}&\frac{z^n}{n!}\sqrt{n!}h_n = \sqrt{4\pi}b\,\times\\&\int_{-\infty}^{\infty}\mathrm{d}y\,K(y)\int_{-\infty}^\infty\mathrm{d}w\, \alpha(y-b(w-w_0))\alpha(w)e^{\sqrt{2}wz-z^2/2}.
\end{split}
\end{equation}
Performing the $w$ integration leaves us with a single integral that is the same as the integral in equation~\eqref{eqn::general_conv_errors} with $t^2=1+b^2$ i.e. $s=1$. We therefore can express $h_n$ in the compact form
\begin{equation}
    h_n = \sqrt{4\pi}\sqrt{\frac{2^{n}}{n!}}\frac{\partial^n}{\partial z^n}\Big[e^{z^2/4}f_1(z)\Big]_{z=0}.
    \label{eqn::general_hn_expression}
\end{equation}
We conclude that if the error-convolved model has an analytic form, the Gauss-Hermite coefficients and error-deconvolved models can also be expressed analytically. The requirement $h_1=h_2=0$ is equivalent to $f_1(z)\sim e^{-z^2/4}(1+qz^3)$ for small $z$ and constant $q$. To match the definition of $h_n$ given in equation~\eqref{eqn::gauss_hermite}, we normalize $h_n$ by $h_0=\sqrt{4\pi}f_1(0)$ such that the normalized $h_0=1$. All subsequent references to $h_n$ are to their normalized versions.

\subsubsection{Model parametrization}

There are two options for how to restrict the kernels (in particular, the parameters $b$ and $w_0$) when modelling: either we choose to fix the mean $\mu=0$ and variance $v=1$, or we fix $h_1=h_2=0$. The first option is analytically simpler as we have shown the moments of $f(w)$ are related to moments of $K(y)$. The latter option follows the parametrization of \cite{vanderMarelFranx1993} who only consider non-zero $h_n$ for $n\geq3$. In general, $h_n$ are complicated functions of the model parameters which do not have analytic roots, so we can only set $h_1=h_2=0$ approximately. We know the two options are equivalent for narrow kernels when $K(y)=\delta(y)$. In the wide $K(y)$ limit, we can use equation~\eqref{eqn::wide_hn} to simplify the mathematics. Our general procedure for setting $h_1\approx h_2\approx0$ is then stitching together the constant variance solution and the wide kernel solution. For symmetric kernels $K(y)$ ($\tilde\mu_n(0)=0$ for odd $n$) setting $w_0=0$ ensures $\mu=0$ and $h_1=0$ for all $b$.

\begin{figure*}
    \centering
    \includegraphics[width=0.9\textwidth]{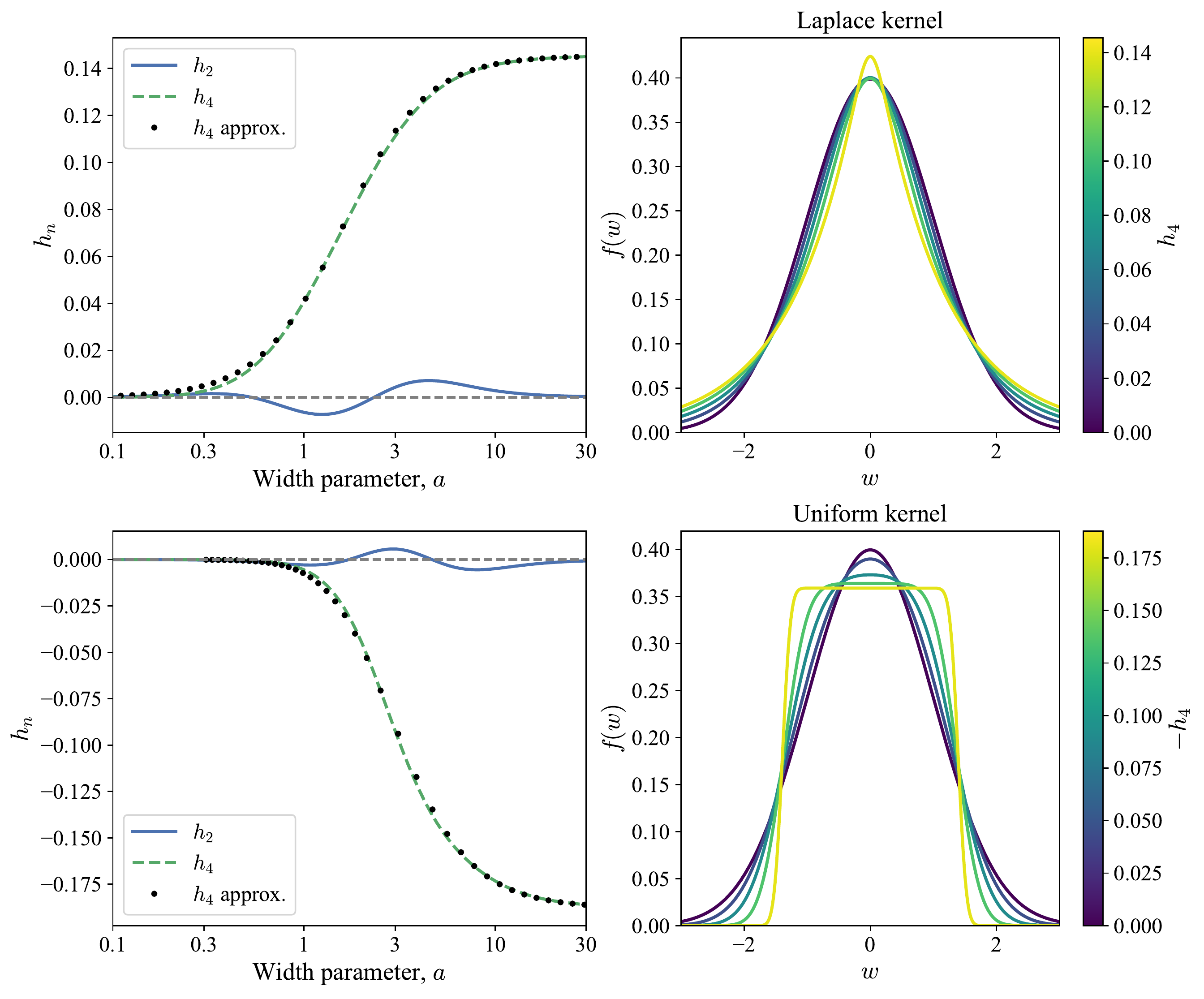}
    \caption{
    Our two families of pdfs with positive (top) and negative (bottom) excess kurtosis (setting skewness or $h_3$ to zero). The left panels show the 2nd (blue) and 4th (green) Gauss-Hermite coefficients of the models as a function of the model parameter, $a$. The model parameters have been selected to minimise variation of $h_2$. The right panels show the probability density functions in terms of the scaled coordinate $w=(x-V)/\sigma$ for different choices of $h_4$.
    }
    \label{fig:newmodel}
\end{figure*}

\subsubsection{Half-kernels}\label{sec::halfkernels}

The parity of the model $f(w)$ is equal to the parity of the kernel $K(y)$. We will find it useful to build arbitrary kernels from half-kernels
\begin{equation}
    K(y) = 
    \begin{cases}
        K_+(y)&y\geq0\\
        K_-(y)&y<0,\\
    \end{cases}
\end{equation}
such that $f(w)=f_+(w)+f_-(w)$. For a positive half-kernel $K_+(y)$, the corresponding negative half-kernel is given by $K_-(y)=K_+(-y)$. Therefore, by the symmetry properties of $f(w)$ and $K(y)$, we have $f_-(w)=f_+(-w)$. Additionally, the characteristic functions will satisfy the symmetry $\phi_{K-}(u)=\phi_{K+}(-u)$. In general, we can stitch together any two half-kernels to build a full kernel. This is a simple way to introduce skewness. There are two possibilities: either we ensure $K_+(0)=K_-(0)$ or $\int_{0}^\infty\mathrm{d}y\,K_+(y)=\int_{-\infty}^0\mathrm{d}y\,K_-(y)=\tfrac{1}{2}$. Here we use the latter choice.

\section{Specific choices of kernel}\label{section::kernels}

With our general theory established, we now turn to some possible choices of kernel $K(y)$. We note here the similarity between our procedure and the method of \cite{LongMurali} for the construction of barred potentials via convolution with a kernel along the bar's major axis. \cite{WilliamsEvans2017} and \cite{McGough2020} showed how flat-topped and cuspy barred potentials could be constructed by this method through convolution with a uniform and Laplace kernel respectively. Here we will follow those choices to construct two families of pdfs.

\subsection{A negative excess kurtosis family -- a uniform kernel}
We first explore a simple symmetric kernel that produces a family of models with negative excess kurtosis. We later expand this model to incorporate skewness. As the models are symmetric $w_0=0$ satisfies both $\mu=0$ and $h_1=0$. Our choice for $K(y)$ is given by a top-hat function
\begin{equation}
K(y) = 
\begin{cases}
\frac{1}{2a},&\mathrm{if}\,|y|<a,\\
0,&\mathrm{otherwise},
\end{cases}
\end{equation}
for $a>0$,
which 
has an error-convolved model given by
\begin{equation}
\begin{split}
    f_s(w)=\frac{b}{2a}\Big[\Phi\Big(\frac{bw+a}{t}\Big)-\Phi\Big(\frac{bw-a}{t}\Big)\Big],
    \label{eqn::uniform_kernel}
\end{split}
\end{equation}
where $t^2=1+b^2s^2$ and $\Phi(x)$ is the cumulative distribution function for the unit Gaussian pdf:
\begin{equation}
    \Phi ( x ) = \frac { 1 } { \sqrt { 2 \pi } } \int _ { - \infty } ^ { x }\mathrm{d}t\,e ^ { - t ^ { 2 } / 2 }.
\end{equation}
The model has zero mean, and the variance and excess kurtosis are given by
\begin{equation}
    v=\frac{1}{b^2}\Big(1+\frac{a^2}{3}\Big), \quad\kappa = -\frac{2a^4}{15}
    \Big(1+\frac{a^2}{3}\Big)^{-2}.
\end{equation}
We observe that there is a simple choice for $b$ such that the variance is unity: $b^2=(1+a^2/3)$.

\begin{figure}
    \centering
    \includegraphics[width=\columnwidth]{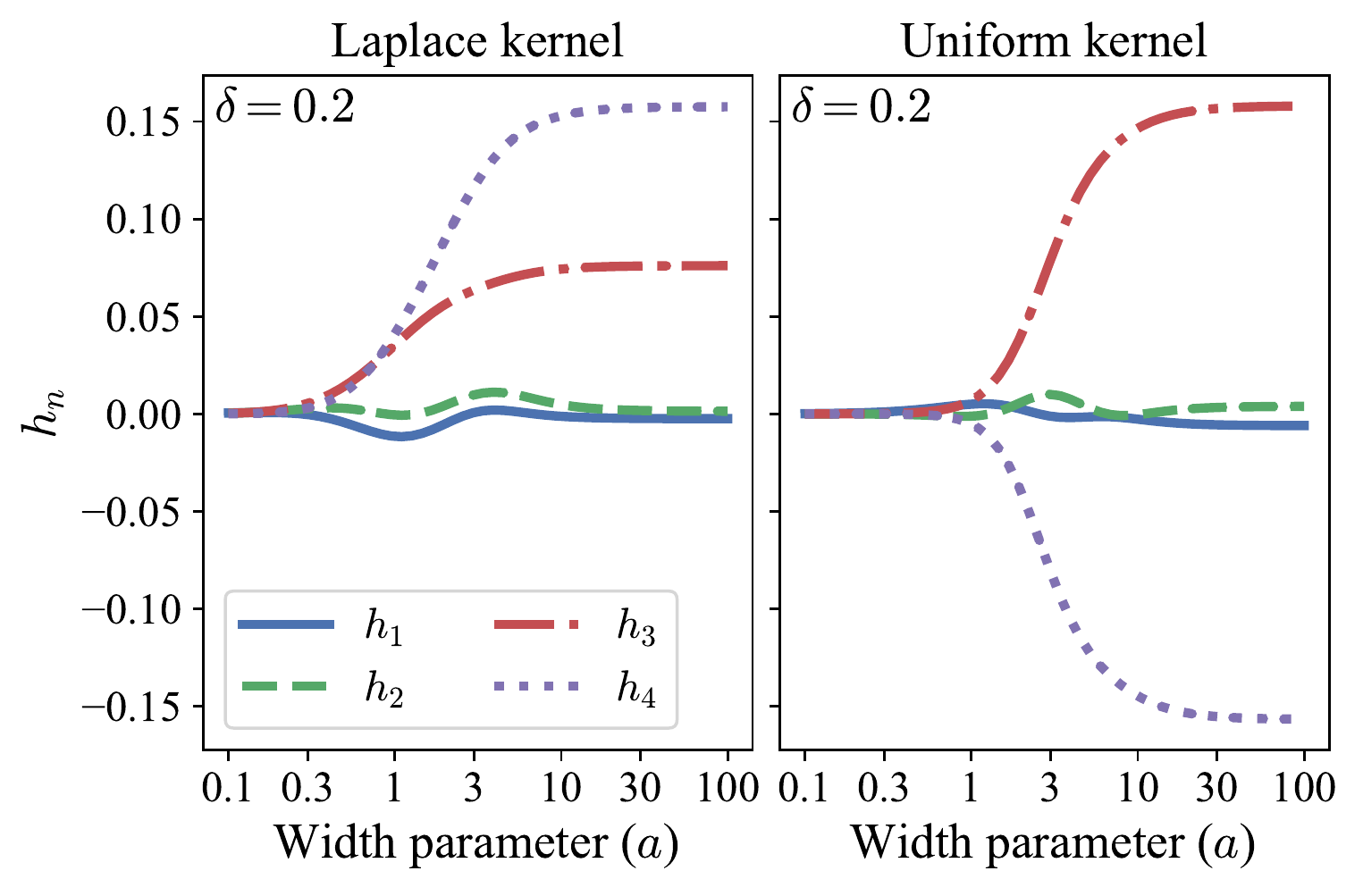}
    \includegraphics[width=\columnwidth]{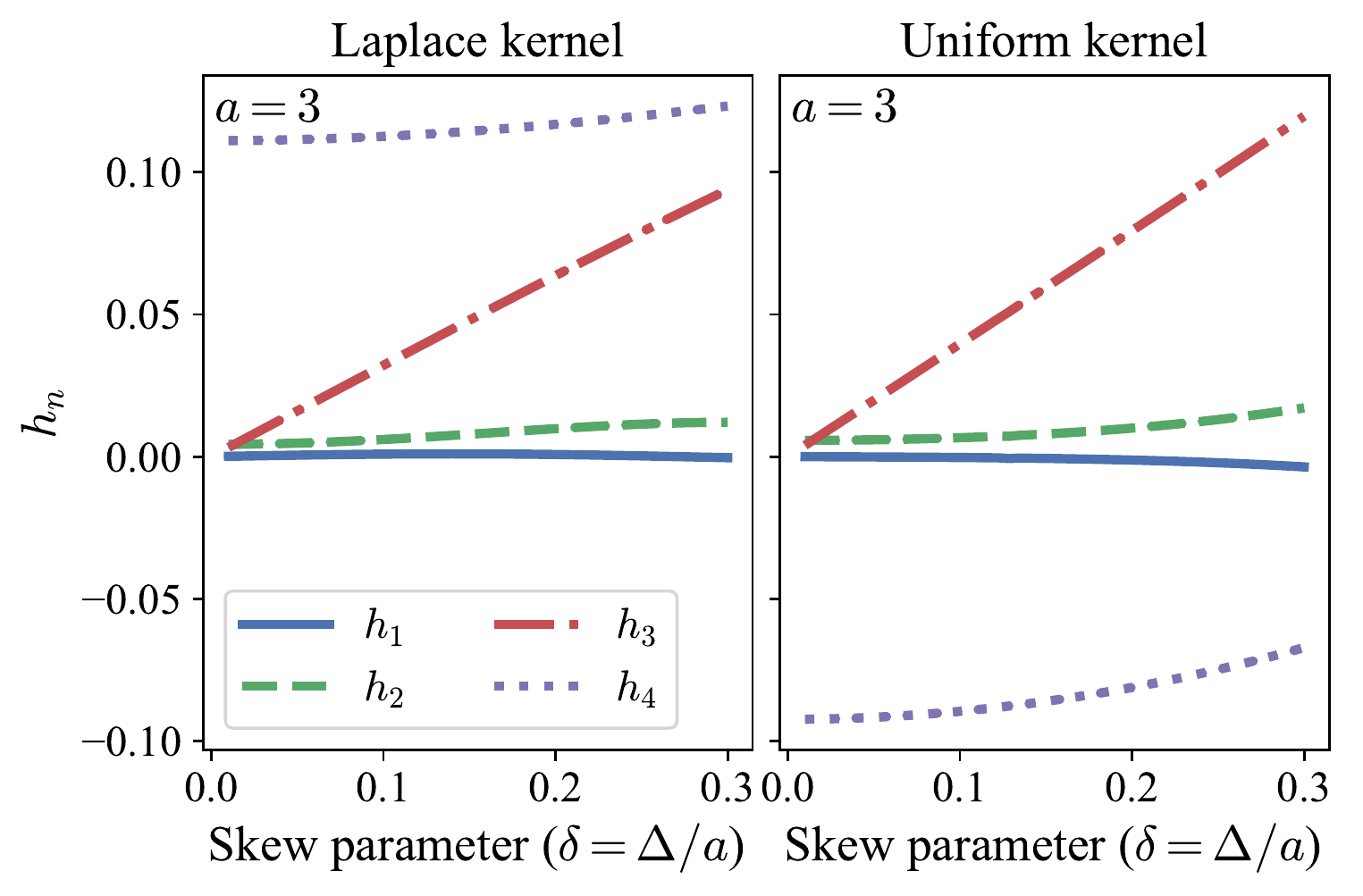}
    \caption{The Gauss-Hermite coefficients, $h_n$, for our two families of models. The left panels correspond to the positive excess kurtosis family with an Laplace kernel and the right to the negative excess kurtosis family with a uniform kernel. The top panels show the variation of $h_n$ with the width parameter $a$ (at fixed $\delta=0.2$) and the bottom variation of $h_n$ with the skewness parameter $\delta=\Delta/a$ (at fixed $a=3$). Note that $h_1\approx0$ and $h_2\approx0$ by design to mirror the Gauss-Hermite series from \protect\cite{vanderMarelFranx1993}.}
    \label{fig:hn_vs_params}
\end{figure}

\subsubsection{Fixing the Gauss-Hermite coefficients}

We now turn to the selection of $b$ such that $h_1\approx h_2\approx0$. As already highlighted, this cannot be done exactly and so only approximate expressions can be derived and choices of fitting functions must be made. For the uniform model, we explicitly detail the procedure we follow, which can be approximately followed for any other choice of model.

In the limit $a\rightarrow\infty$, we require $a/b\rightarrow k$ for constant $k$ such that the variance and $h_2$ remain finite. 
We introduce a general functional dependence of $b$ of the form
\begin{equation}
    b^2 = 1+\Big(\frac{a}{k}\Big)^2,
\end{equation}
which encompasses the low (unity variance) and high $a$ limits for different choices of $k$. To recover the constant variance choice, we can set $k=k_0=\sqrt{3}$. Using equations~\eqref{eqn::wide_hn} and~\eqref{eqn::general_hn_expression}, we determine the normalized $h_2$ and $h_4$ for large $a$ as
\begin{equation}
\begin{split}
\lim_{\substack{a\rightarrow\infty\\a/b\rightarrow k}}h_2&=\sqrt{\frac{1}{2}}-\frac{2}{\sqrt{\pi}}ke^{-k^2/2}\Big/\mathrm{erf}\Big(\frac{k}{\sqrt{2}}\Big),\\ 
\lim_{\substack{a\rightarrow\infty\\a/b\rightarrow k}}h_4&=\sqrt{\frac{3}{8}}-\frac{2}{\sqrt{3\pi}}k^3e^{-k^2/2}\Big/\mathrm{erf}\Big(\frac{k}{\sqrt{2}}\Big).
\end{split}
\end{equation}
We observe that there is a choice of $k=k_\infty$ which sets $\lim_{a\rightarrow\infty}h_2=0$ given by $k_\infty=1.399985\dots$, amazingly close to $7/5$. With this choice, $\lim_{a\rightarrow\infty}h_4=-0.187777\dots\equiv h_{4,\infty}$.
Therefore, to replicate the modelling with a Gauss-Hermite series which imposes the requirement that $h_2=0$ we can set $k=k_\infty$. This choice produces $|h_2|\lesssim0.05$ everywhere. To improve this, we make $k$ a weak function of $a$ which reproduces the constant variance limit $k=k_0$ for small $a$ as
\begin{equation}
k(a)=k_0-(k_0-k_\infty)\tanh(a/a_0),
\end{equation}
where setting $a_0=3.3$ minimizes the variation of $h_2$ to $|h_2|\lesssim0.006$. We show this set of models in the bottom panels of Fig.~\ref{fig:newmodel}. We find that a good approximation for $a$ given $h_4$ is
\begin{equation}
    a=a_{h4}\Big(\sqrt{\frac{h_{4,\infty}}{h_4}}-1\Big)^{-1/2},
    \label{eqn::a_fnh}
\end{equation}
where $a_{h4}=2$. This fitting function is shown in Fig.~\ref{fig:newmodel}
. In this way, we can parameterize our models in terms of $h_4$.

\begin{table*}
    \caption{Key formulae for the two families of distributions introduced in this paper. $f_s(w)$ is the probability density function of the scaled and shifted coordinate $w=(x-V)/\sigma$ convolved with observational uncertainties $\sigma_e$ such that $s=\sigma_e/\sigma$. $\Phi(x)$ is the cumulative of the unit normal distribution. Further generic definitions are $w'=w-w_0$, $t=1+b^2s^2$, $a_+=a+\Delta$, $a_-=a-\Delta$, $\delta=\Delta/a$, $b^2 = 1+(a/k)^2$ and $k = k_0 - (k_0-k_\infty)\tanh(a/a_0)$. The parameters $a$, $b$ and $\Delta$ are chosen as per the `Auxiliary Functions' and `Parameters' columns to ensure $h_1\approx h_2\approx0$ and to match a given $h_3$ and $h_4$. For a given $h_3$ and $h_4$, the quadratic for $\delta$ is solved, $a$ is computed using $h_{4,\mathrm{max}}(\delta)$ and finally $b$ and $w_0$ are found. Code for the computation of all formulae is provided at \href{https://github.com/jls713/gh_alternative}{https://github.com/jls713/gh\_alternative}.
       }
    \centering
    \begin{tabular}{l|l|l|llllll}
         \hline
         Name&$f_s(w)$&Auxiliary Functions&
         \multicolumn{6}{l}{Parameters}\\%&&&&  \\
        &&& $h_{4,\infty}$&$a_0$&$a_{h3}$&$a_{h4}$&$k_0$&$k_{\infty,0}$\\
         \hline
         Uniform&
         $
        \!\begin{aligned}
         \frac{b}{2a_+a_-}\Big[a_+\Phi\Big(\frac{bw'+a_-}{t}\Big)-a_-\Phi&\Big(\frac{bw'-a_+}{t}\Big)\\&-2\Delta\Phi\Big(\frac{bw'}{t}\Big)\Big]
        \end{aligned}
         $&
         $\def\arraystretch{1.6}
        \begin{array}{l}
        h_{4,\mathrm{max}}=(1-4.3\delta^2)h_{4,\infty}\\ h_{3,\mathrm{max}}=0.82\delta\\
         a = a_{hi}\Big(\sqrt{\frac{h_{i,\mathrm{max}}}{h_i}}-1\Big)^{-\tfrac{1}{2}}\\
        k_\infty=k_{\infty,0}\sqrt{1+\delta^2+3\delta^4}\\
        w_0 = -\frac{\Delta}{2b}+\frac{\Delta}{3b}\tanh\Big(\frac{a}{a_0}\Big)
        \end{array}
        $
        &
        % $0.2$
        $-0.1878$
        &$3.3$&$2$&$2$&$\sqrt{3}$&$\frac{7}{5}$
        \\
         \hline
        Laplace&
        $\!\begin{aligned}
&\frac{b}{4a_+}\exp\Big(\frac{t^2-2a_+bw'}{2a_+^2}\Big)\mathrm{erfc}\Big(\frac{t^2-a_+bw'}{\sqrt{2}ta_+}\Big)
    \\&+\frac{b}{4a_-}\exp\Big(\frac{t^2+2a_-bw'}{2a_-^2}\Big)\mathrm{erfc}\Big(\frac{t^2+a_-bw'}{\sqrt{2}ta_-}\Big)
\end{aligned}$&
    $
    \def\arraystretch{1.6}
    \begin{array}{l}
         h_{4,\mathrm{max}}=(1+2\delta^2)h_{4,\infty}\\
         h_{3,\mathrm{max}}=0.37\delta\\
         a = a_{hi}\Big(\frac{h_{i,\mathrm{max}}}{h_i}-1\Big)^{-\tfrac{1}{2}}\\
        k_\infty=k_{\infty,0}\sqrt{1+3\delta^2}\\
        w_0 = -\frac{\Delta}{b}+\frac{8\Delta}{7b}\tanh\Big(\frac{5a}{4a_0}\Big)\\
        \end{array}
        $
        &
        % $0.13$
        $0.1454$
        &$2.25$&$1.1$&$1.6$&$\frac{1}{\sqrt{2}}$&$1.08$\\
         \hline
    \end{tabular}
    \label{tab:keyformulae}
\end{table*}

\subsubsection{Incorporating skewness}

Following the discussion in Section~\ref{sec::halfkernels}, a generalization of this model which incorporates some skewness is found by using a different uniform distribution for positive and negative $y$ as
\begin{equation}
K(y) = 
\begin{cases}
\frac{1}{2a_+},&\mathrm{if}\,0<y<a_+,\\
\frac{1}{2a_-},&\mathrm{if}\,-a_-<y<0,\\
0,&\mathrm{otherwise},
\end{cases}
\end{equation}
for $a_+>0$ and $a_->0$
which produces 
an error-convolved model of the form
\begin{equation}
\begin{split}
    f_s(w)
    =\frac{b}{2a_+a_-}\Big[a_+\Phi\Big(\frac{bw'+a_-}{t}\Big)-a_-\Phi\Big(\frac{bw'-a_+}{t}\Big)-2\Delta\Phi\Big(\frac{bw'}{t}\Big)\Big],
\end{split}
\label{eqn::uniform_skewness}
\end{equation}
where $w'=w-w_0$, $a=\tfrac{1}{2}(a_++a_-)$ and $\Delta=\tfrac{1}{2}(a_+-a_-)<a$. We describe numerical calculation of $f_s(w)$ in Appendix~\ref{appendix::numerical_implementation}. In Table~\ref{table::cumulants}, we provide the cumulants for this model. As expected, $\Delta$ controls the skewness of the distribution, which we parametrize in a scale-free way using $\delta=\Delta/a$.

We use equation~\eqref{eqn::wide_hn} to find expressions for $h_1$ and $h_2$ in the limit of large $a$ and $a/b=k$. Adopting the choice of $k$ for $\Delta=0$, $k=k_\infty=7/5$, we find $w_0\approx-\Delta/6b$ approximately sets $h_1=0$. We also find that $k_\infty$ must be made a function of $\delta=\Delta/a$ to retain $h_2=0$. With these choices, we set
\begin{equation}
\begin{split}
    k_\infty &= \frac{7}{5}\sqrt{1+\delta^2+3\delta^4},\\
    w_0 &= -\frac{\Delta}{2b}+\frac{\Delta}{3b}\tanh(a/a_0).
\end{split}
\end{equation}
where $\delta=\Delta/a$. We find that the maximum $|h_4|$ is $h_{4,\mathrm{max}}\approx(1-4.3\delta^2)h_{4,\infty}$ and that the maximum $h_3$ is $|h_{3,\mathrm{max}}|\approx0.82|\delta|$. We adopt the same functional form for $h_3(a)$ as $h_4(a)$, which means $h_3/h_4 = h_{3,\mathrm{max}}/h_{4,\mathrm{max}}$. For a choice of $h_3$ and $h_4$, we first find $\delta$ by solving a quadratic. We then obtain $a$ from equation~\eqref{eqn::a_fnh}. The full set of parameters describing the model are given in Table~\ref{tab:keyformulae}.

In Fig.~\ref{fig:hn_vs_params}, we show the variation of $h_n$ for this model as a function of $a$ and $\delta$. We see how our choices have ensured $h_1\approx0$ and $h_2\approx0$. $\delta$ approximately controls $h_3$ at fixed $a$, and $a$ controls $h_4$ although there is some variation of $h_4$ with $\delta$.

\subsection{A positive excess kurtosis family -- a Laplace kernel}

An extension to positive excess kurtosis can be constructed using a Laplace kernel. The exponential naturally produces broader tails than the Gaussian. The choice of kernel is
\begin{equation}
K(y) = 
\begin{cases}
\frac{1}{2a_+}e^{-y/a_+},&\mathrm{if}\,y>0,\\
\frac{1}{2a_-}e^{y/a_-},&\mathrm{if}\,y<0,
\end{cases}
\end{equation}
for $a_+>0$ and $a_->0$,
which produces an error-convolved distribution of the form
\begin{equation}
\begin{split}
    f_s(w) &= \frac{b}{4a_+}\exp\Big(\frac{t^2-2a_+bw'}{2a_+^2}\Big)\mathrm{erfc}\Big(\frac{t^2-a_+bw'}{\sqrt{2}ta_+}\Big)
    \\&+\frac{b}{4a_-}\exp\Big(\frac{t^2+2a_-bw'}{2a_-^2}\Big)\mathrm{erfc}\Big(\frac{t^2+a_-bw'}{\sqrt{2}ta_-}\Big),
    \label{eqn::laplace}
\end{split}
\end{equation}
where $w'=w-w_0$.
This model is a sum of exponentially modified Gaussian distributions. Note that the products of exponentials and error functions require careful numerics when the arguments of the exponentials are large and positive. We discuss numerical implementation details in Appendix~\ref{appendix::numerical_implementation}.

In Table~\ref{table::cumulants}, we provide the cumulants for this model. We follow a similar procedure to the uniform kernel case for restricting the parameters $b$ and $w_0$ such that $h_1\approx h_2\approx0$. One key difference is that our choice of parametrization of $h_i(a)$ is
\begin{equation}
    a=a_{hi}\Big(\frac{h_{i,\mathrm{max}}}{h_i}-1\Big)^{-1/2},
    \label{eqn::a_fnh_exp}
\end{equation}
with different scale parameters $a_{hi}$ for $h_3$ and $h_4$.
We show the accuracy of this approximation in Fig.~\ref{fig:newmodel} along with the pdfs for $h_3=0$.
The full model parametrization is given in Table~\ref{tab:keyformulae}.

In Fig.~\ref{fig:hn_vs_params}, we show the variation of $h_n$ for this model as a function of $a$ and $\delta$ alongside the equivalent for the uniform kernel. We see our choices have made $h_1\approx0$ and $h_2\approx0$. $h_3$ is controlled by $\delta$ at fixed $a$, and varying $a$ varies both $h_3$ and $h_4$ together.

\subsection{Further choices of kernel}

The general procedure we have employed for the uniform and Laplace kernels is applicable for a broader choice of kernels. The sole requirement is that the convolution of the kernel with a Gaussian is analytic. We have considered a number of other kernels with both positive and negative excess kurtosis. In Table~\ref{table::kernels}, we detail a number of half-kernels $K_+(y)$ along with their corresponding error-convolved distributions $f_{s+}(w)$. Along with the uniform and Laplace kernels, we give expressions for a raised cosine distribution, a Gaussian distribution and a gamma distribution. In Table~\ref{table::cumulants}, we give the cumulants produced by stitching together two half-kernels with different scales ($a_\pm$). The raised cosine distribution produces a negative excess kurtosis family ($\kappa=-0.59$) and can be asymmetrized in the same way as the Laplace family. The Gaussian family produces a family of positive excess kurtosis but non-zero excess kurtosis is always accompanied by skewness. The gamma distribution, which is only defined for positive $y$, produces a positive excess kurtosis family.
We note that the form of the models we have introduced encompasses the well-known Voigt profile which is formed from the convolution of a Cauchy distribution with a Gaussian. It is regularly used in spectroscopy, and can be computed using the real part of the Faddeeva function.

\begin{figure}
    \centering
    \includegraphics[width=\columnwidth]{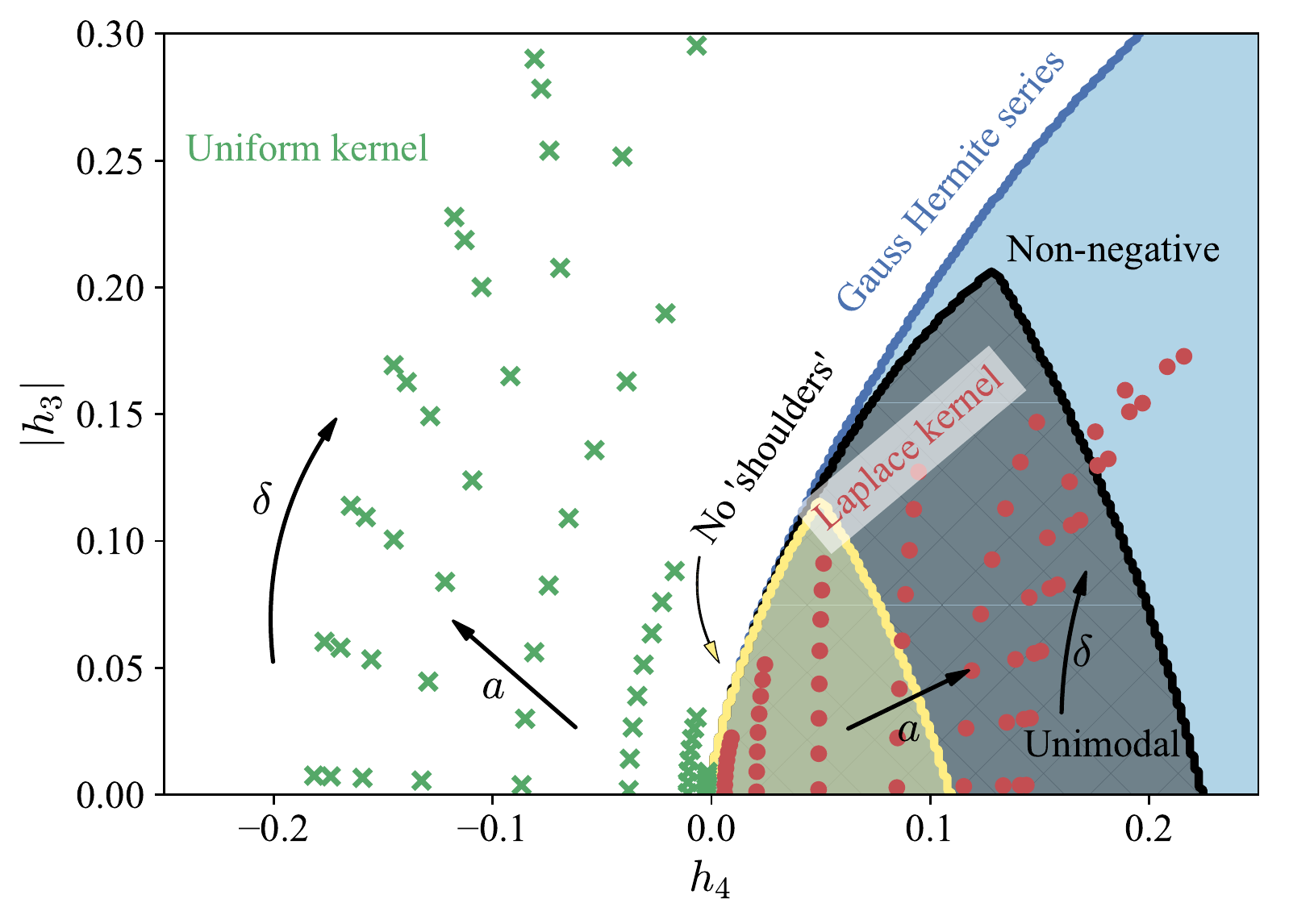}
    \caption{The space of Gauss-Hermite coefficients accessible by our new models: red points show the positive excess kurtosis family formed by the Laplace kernel whilst green crosses show the negative excess kurtosis family formed from the uniform kernel. The parameters $a$ and $\delta=\Delta/a$ act approximately like polar coordinates in this space. The points are spaced logarithmically in $a$ and linearly in $\delta$ (up to a maximum of $\delta=0.5$ as this is the range over which our approximations are valid). A Gauss-Hermite series can be constructed for all parts of the figure but only in the blue region are the series non-negative everywhere, only within the black hatched region are they unimodal and non-negative, and only within the yellow region do they have a fixed sign of the curvature in the tails.}
    \label{fig:parameter_space}
\end{figure}

\subsection{Comparison with the Gauss-Hermite series}

We close our investigation of the new families by comparison to the Gauss-Hermite series. 
% \citep{vanderMarelFranx1993}. 
In Fig.~\ref{fig:parameter_space}, we show the range of $h_3$ and $h_4$ accessible by our two families of models. We show for comparison the region of $(h_3,h_4)$ space in which the Gauss-Hermite series are non-negative, non-negative and unimodal, and non-negative, unimodal and with `no shoulders' i.e. the sign of the curvature of the tails doesn't change. In this space, the parameters $(a,\delta)$ behave approximately as polar coordinates. The regions of accessible $(h_3,h_4)$ form wedges highlighting a limitation of our models that the range of available $h_3$ depends upon $h_4$ and vice-versa. However, the same is true of the non-negative group of Gauss-Hermite models. The unimodal, non-negative Gauss-Hermite series essentially encompasses the range of models accessible by the Laplace kernel, which suggests the Gauss-Hermite is preferable for modelling positive excess kurtosis. However, the advantage of our new family is that non-negativity is built in and does not need to be checked numerically for each model. Additionally, all of our new positive excess kurtosis models have a constant sign for the curvature in the tails, whilst for smaller positive $h_4$ values than admitted by the new family the Gauss-Hermite can develop wings or shoulders.

\begin{figure}
    \centering
    \includegraphics[width=\columnwidth]{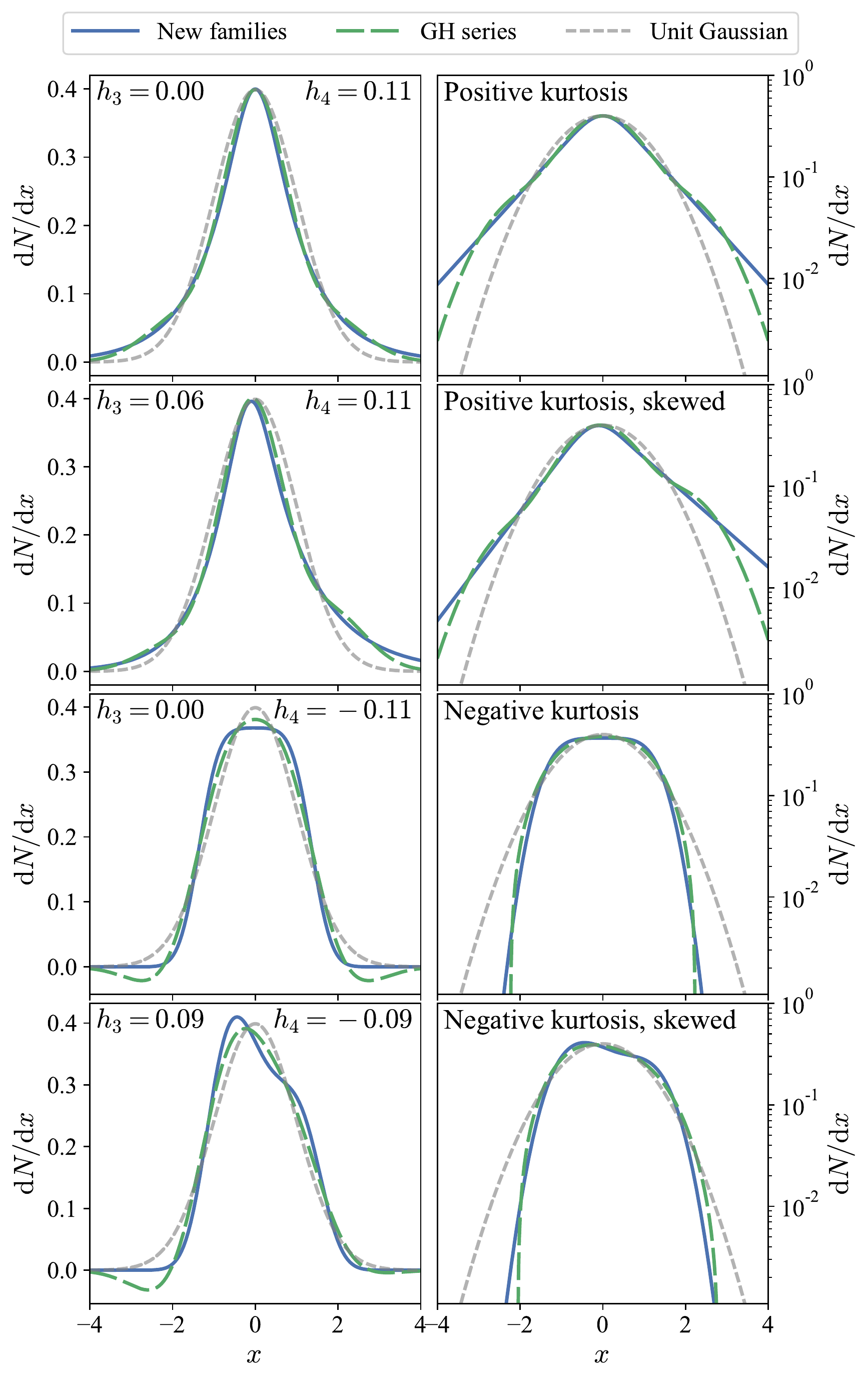}
    \caption{Comparison between the new models and the Gauss-Hermite series for four different choices of $h_3$ and $h_4$ (labelled in each row). Each set of panels show the new models in blue solid, the equivalent Gauss-Hermite series in green long-dashed and a reference unit Gaussian in light grey short-dashed. The left panels have a linear $y$ axis and right logarithmic. The Gauss-Hermite series models are normalized to have unit weight over the region of positivity. The two sets of models match well but for negative excess kurtosis the negative regions of the Gauss-Hermite series are avoided.}
    \label{fig:ghcomparison}
\end{figure}

In Fig.~\ref{fig:ghcomparison}, we show the Gauss-Hermite series and the new family of models for four choices of $h_3$ and $h_4$. There is a great deal of similarity between the two different models although importantly the new models avoid the regions of negative probability for negative excess kurtosis. We also note that the new family of positive excess kurtosis models has stronger tails than the Gauss-Hermite series showing the presence of higher order Gauss-Hermite coefficients in their expansion. Another generic feature of our models that doesn't appear to arise in the Gauss-Hermite series expansion is the presence of a `shoulder' for skewed negative excess kurtosis models: a feature due to our method of stitching together half-kernels.

\section{Application to dwarf spheroidal data}\label{section::application}

As an illustration of our approach, we will demonstrate its use on line-of-sight velocity data from dwarf spheroidal galaxies of the Milky Way. Dwarf spheroidal galaxies are gas-poor elliptical low stellar mass systems \citep{Mateo1998}. Within $\Lambda$ cold dark matter theory these objects are anticipated to be highly dark-matter dominated. This appears to be borne out by line-of-sight velocity measurements for the dwarf spheroidal galaxies of the Local Group and so make these objects ideal laboratories for investigating the properties of dark matter. The density profile of dark matter within these objects has received considerable attention, in particular the question of whether the central regions are cored or cuspy. The profile shape is a probe of both the effects of baryonic feedback on cold dark matter, or the properties of other dark matter candidates (such as self-interacting dark matter). However, pinning down the density profile is not a simple task due to the well-known mass-anisotropy degeneracy. Solutions to this include the use of proper motions \citep{Massari2018} or performing analyses on different sub-populations within the dwarf spheroidals \citep{WalkerPenarrubia2011,AE12ds}. However, it has also been suggested that the velocity distribution contains sufficient information to break the mass-anisotropy degeneracy in its higher order moments, in particular the kurtosis \citep[e.g.][]{Lokas2002,Richardson2013,Read2019}. As highlighted by \cite{vanderMarelFranx1993}, kurtosis is a measure of orbital anisotropy with positive excess kurtosis associated with radial orbits. Here we will investigate how our new models can be used to constrain the higher order velocity moments from dwarf spheroidal data. We will focus on the specific example of the Fornax dSph as it has the largest line-of-sight velocity dataset.

\subsection{Dynamical models of the Fornax dSph}

\begin{figure*}
    \centering
    \includegraphics[width=\textwidth]{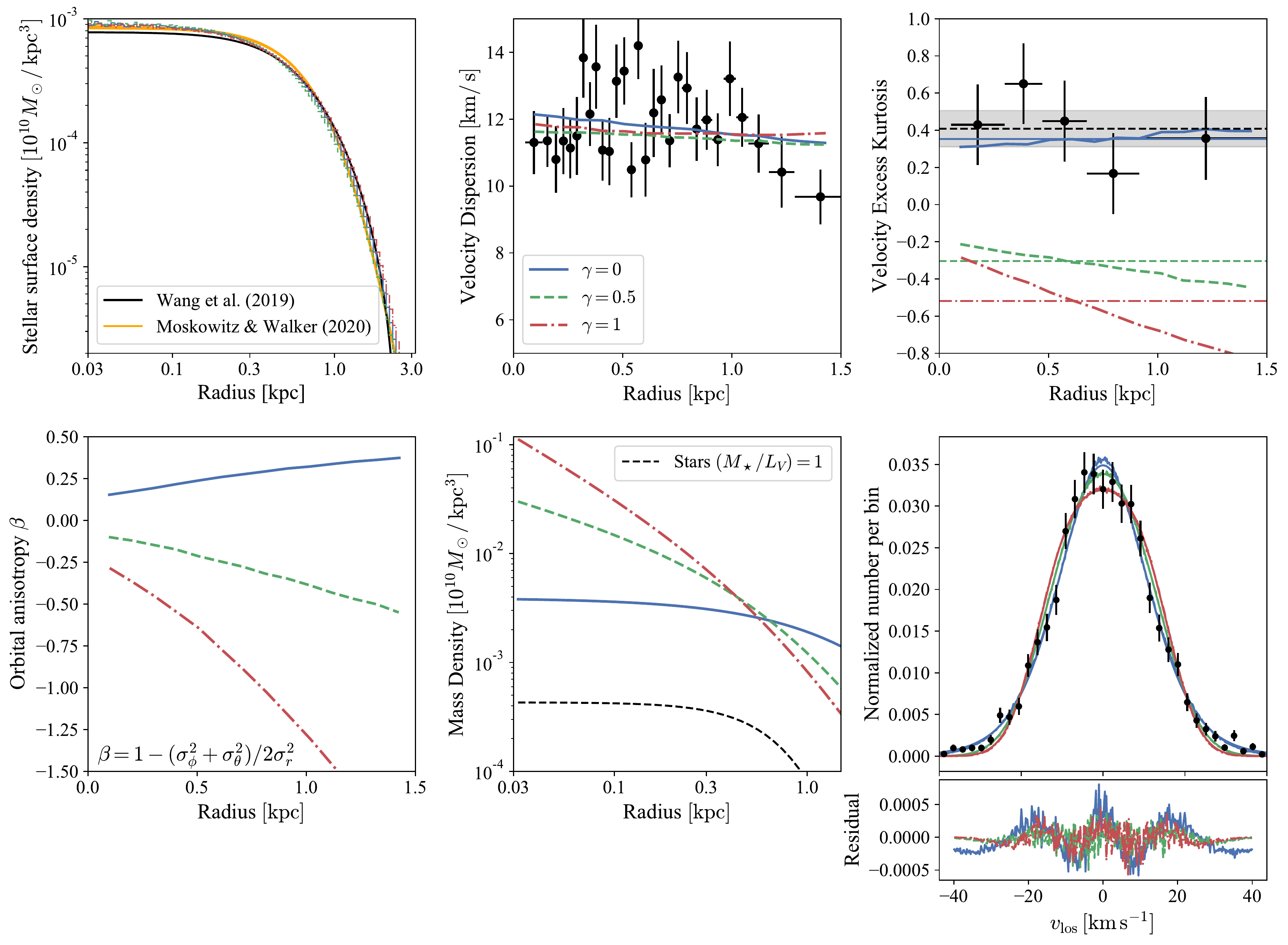}
    \caption{Dynamical models of the Fornax dwarf spheroidal galaxy. Data are always shown in black (statistics are weighted using membership probabilities from subsection~\ref{sec::kurtfit}, $x$ errorbars show the bin size and $y$ errorbars are from Poisson only, e.g. equation~\eqref{eqn::standard_errors}, so do not consider measurement uncertainties, although the dispersion is deconvolved by the median uncertainty) and the three models are blue solid (cored, $\gamma=0$), green dashed (weakly cusped, $\gamma=-0.5$) and red dash-dot (strongly cusped, $\gamma=-1$). 
    The top left panel shows the density profiles as a function of radius. For the data, we always use the circularized radius $\sqrt{1-\epsilon}(x^2+y^2/(1-\epsilon)^2)^{1/2}$ for flattening $\epsilon=0.31$ to compare to our spherical models. We also display the posterior distribution of the density fits to DES data from \protect\cite{Moskowitz2020} and the King fit to DES data from \protect\cite{Wang2019} scaled to have $L_V=1.4\times10^7\,L_\odot$ with $(M_\star/L_V)=1$. The top middle panel shows the velocity dispersion and top right the excess kurtosis of the velocity distribution. The horizontal lines are the mean excess kurtosis of the models. The black band is the median and $\pm1\sigma$ excess kurtosis of the model. In the bottom row we show the orbital anisotropy (left) and the mass density (middle) along with stellar density. In the bottom right, we show the model velocity distributions (binned histograms) along with fits using the pdfs in this work. The Fornax data are also shown. The models do an excellent job of capturing the kurtosis of the dynamical models, so much so that they are barely distinguishable in the upper panel, but the deviations are shown in the lower panel of residuals. }
    \label{fig:fornax}
\end{figure*}

We take the spectroscopic sample of $2633$ stars
presented in \cite{Walker2009}, who provide line-of-sight velocities with uncertainties for all stars and a metallicity indicator, a pseudo-equivalent width measurement of the Mg triplet, along with its uncertainty for $459$ stars.
We convert the on-sky positions into projected circularized radius $R_c=\sqrt{1-\epsilon}(x^2+y^2/(1-\epsilon)^2)^{1/2}$ where $x/y$ are the positions along the major/minor axis of Fornax with respect to the dwarf centre. We take the centre of Fornax as $(\alpha,\delta)=(2\mathrm{h}39\mathrm{m}53\mathrm{s},-34^\circ30'32'')$, the position angle as $42.2\,\mathrm{deg}$ and the ellipticity of $\epsilon=0.31$ from the King profile fit of  \cite{Wang2019}. Some targets were observed multiple times by \cite{Walker2009} from which an inverse-variance-weighted mean line-of-sight velocity and its error are estimated. For the remainder of the stars we use the single velocity measurement along with its error. \cite{Pascale2018} discuss the effect of undetected binarity on the Fornax velocity distributions by comparing samples with and without binaries as determined by repeat observations. They find the distributions are very similar (less than $4\percent$ probability the distributions are drawn from different samples) so we conclude binarity is unimportant and do not attempt to correct for it. We correct for the velocity gradient across the dwarf produced by perspective rotation from the bulk proper motion using the equations in \cite{Walker2008} and the proper motion measurements from \cite{HelmiGaiaDR2}. 

In addition to the spectroscopic data, we use the Fornax surface density profile measurements from Dark Energy Survey Data from \cite{Wang2019} and \cite{Moskowitz2020}. \cite{Wang2019} provide a best-fitting King profile to the data although acknowledge all the analytic forms they consider do not exactly match the data. \cite{Moskowitz2020} provide posterior draws from the parameters of a three-component Plummer-like profile, but with a steeper $\rho\propto r^{-9}$ fall-off. We assume a distance to Fornax of $138\,\mathrm{kpc}$ \citep{Mateo1998,Rizzi2007} although our results are not particularly sensitive to this choice. 

\begin{figure*}
    \centering
    \includegraphics[width=\textwidth]{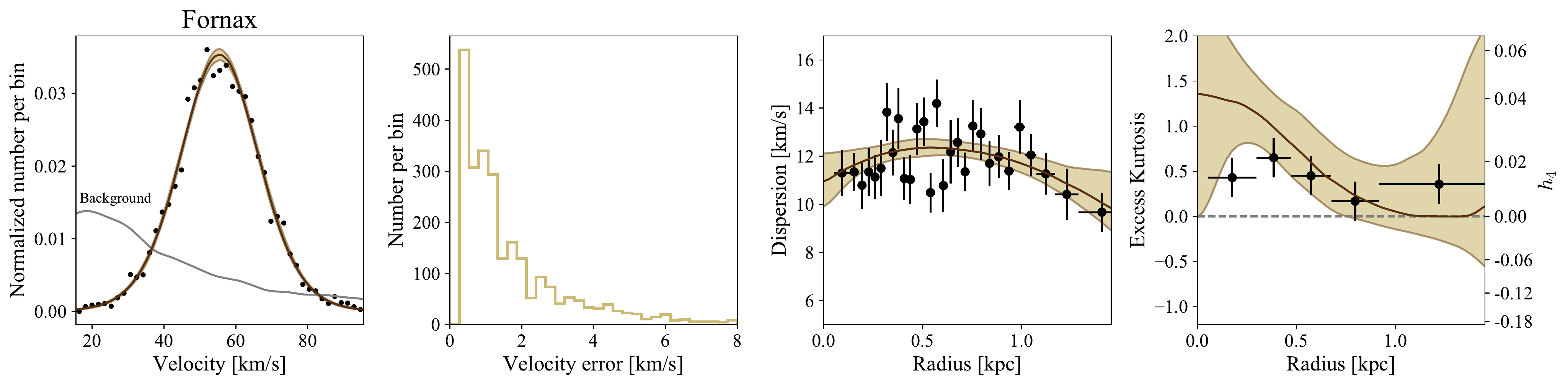}
    \includegraphics[width=\textwidth]{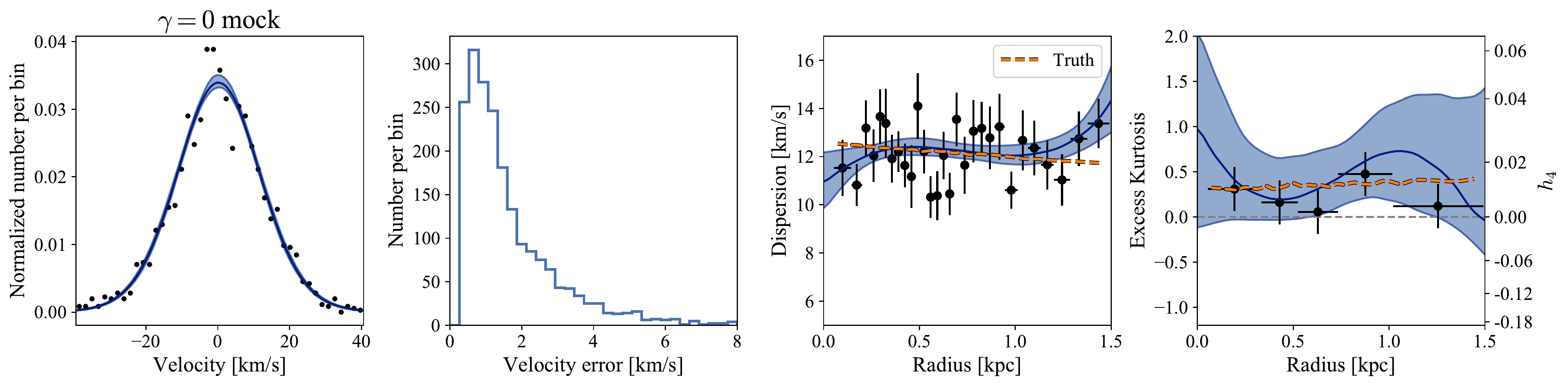}
    \includegraphics[width=\textwidth]{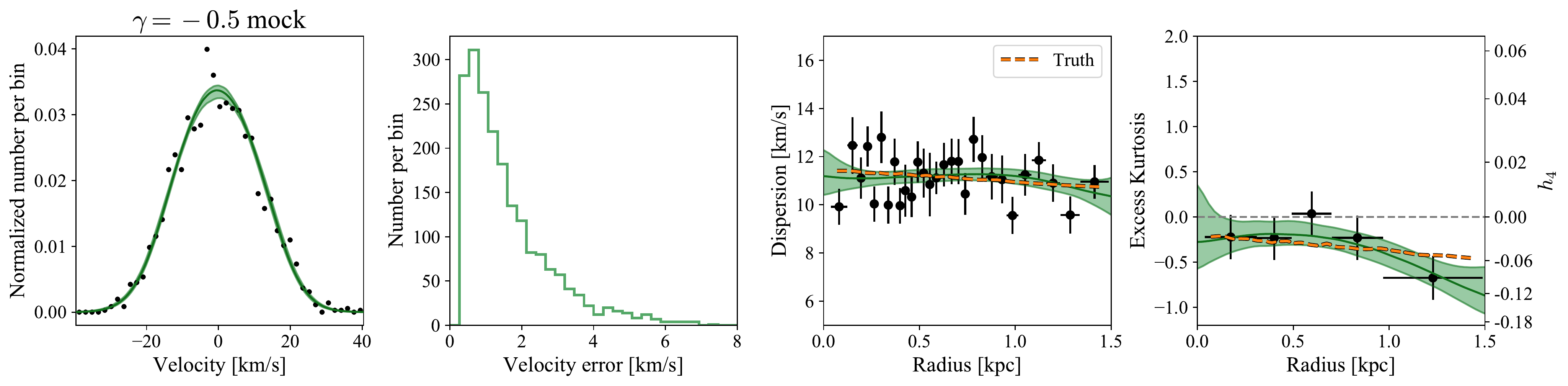}
    \includegraphics[width=\textwidth]{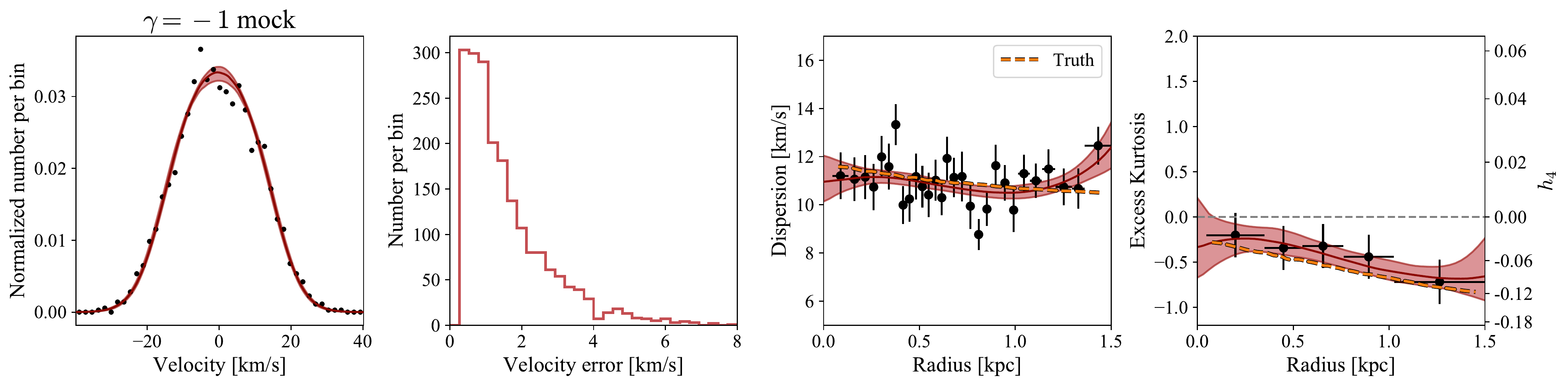}
    \caption{Fits of our family of pdfs to the Fornax data (top row) and three mock Fornax datasets (bottom three rows). The left column shows the velocity distribution of the data (mocks) in black and the median with $\pm1\sigma$ fits of the models. In the top panel we also show the background Galaxia \protect\citep{Sharma2010} model used. The second column shows the distribution of uncertainties. The third and fourth columns show the (membership-weighted) dispersion and excess kurtosis data (black with errorbars showing the bin size in $x$ and Poisson errors in $y$) and fits along with the truth for the mocks in orange.}
    \label{fig:fornax_fit}
\end{figure*}

For constructing models of Fornax, we use the \textsc{Agama} dynamical modelling framework from \cite{Vasiliev2019}. Such an approach has also been taken by \cite{Pascale2018} in fitting density and velocity data on Fornax. To model the stellar distribution, we use the spheroidal distribution functions similar to those developed in work by \cite{WilliamsEvans2015} and \cite{Posti2015}. They take the form
\begin{equation}
    \begin{split}
f(\bs{J}) &= 
\frac{f_0}{(2\pi)^3}
g(\bs{J})^{-\Gamma}
\exp\bigg[-\left(\frac{g(\bs{J})}{J_\mathrm{cutoff}}\right)^\zeta\bigg],\\
g(\bs{J}) &\equiv g_r J_r + \tfrac{1}{2}(3-g_r)(J_z\, + |J_\phi|), 
\end{split}
\end{equation}
The linear combination of the actions $g(\bs{J})$ is set to produce approximately spherical models and $g_r$ controls the degree of radial anisotropy. The models have an exponential cutoff at scale $J_\mathrm{cutoff}$ with the option of a flexible inner slope controlled by $\Gamma$. For the dark matter, we consider the simple family of density profiles
\begin{equation}
    \rho_\mathrm{DM}(r)=\rho_0 (r/r_s)^{-\gamma}(1+r/r_s)^{-3.2}.
\end{equation}
For the potential from the stars, we fit the projection of a 3d power-law of the form $\rho_{\star,0}(1+(r/r_\star)^2)^{-9/2}$ to the \cite{Moskowitz2020} density profile and assume a $V$-band stellar mass-to-light ratio of $1$ and $L_V=1.4\times10^7L_\odot$ \citep{Irwin1995}.

We use three values of the inner dark-matter density slope $\gamma=(0,0.5,1)$ and by-hand approximately match the density profile and the velocity dispersion profile of Fornax as shown in Fig.~\ref{fig:fornax}. The chosen parameters are given in Table~\ref{tab:parameters_dsph}. As our models are spherical, we always compare to the circularized radius $R_c$ distributions for the data.
We also opt to weight the moments calculated from the data by membership probabilities computed in the next subsection and only consider stars within $55\,\mathrm{km\,s}^{-1}$ of the mean velocity of Fornax (approximately $5$ times the dispersion of Fornax).

\begin{table}
    \caption{Parameters of our action-based Fornax models. The models differ primarily in their inner dark matter slope $\gamma$. $\rho_0$ and $r_s$ are the normalization and scale radius of the dark matter halo with units $10^{10}M_\odot/\,\mathrm{kpc}^3$ and $\mathrm{kpc}$. $f_0$ is the normalization of the stellar distribution (with units $10^{10}M_\odot/(\mathrm{kpc\,km\,s}^{-1})^{3-\Gamma}$). $\Gamma$ controls the central density slope of the stars, $J_\mathrm{cutoff}$ (with units $\mathrm{kpc\,km\,s}^{-1}$) and $\zeta$ govern the location and strength of the stellar exponential break, and $g_r$ the radial anisotropy of the stars.}
    \centering
    \begin{tabular}{cccccccc}
    $\gamma$&$\rho_0$&$r_s$&$f_0$&$\Gamma$&$J_\mathrm{cutoff}$&$g_r$&$\zeta$\\
    \hline
$0.0$&$0.0039$&$4.0$&$2\times10^{-5}$&$0.0$&$1.3$&$1.2$&$0.6$\\
$0.5$&$0.005$&$1.3$&$7.6\times10^{-6}$&$0.0$&$3.2$&$1.6$&$0.8$\\
$1.0$&$0.0038$&$1.0$&$3\times10^{-6}$&$0.2$&$8.6$&$1.9$&$1.4$\\
\hline
    \end{tabular}
    \label{tab:parameters_dsph}
\end{table}

We observe that the three models have different anisotropy (the cuspy model is tangentially-biased whilst the cored model is more radially-biased) which correspond to differences in the excess kurtosis. Although all three models have very similar tracer density profiles and velocity dispersion profiles, the mass density distributions and hence anisotropy profiles are very different \citep[e.g.,][]{BinneyMamon1982,Evans2009}. From this admittedly simplistic fitting approach, it seems the kurtosis of the data favours more cored models of Fornax, a conclusion also reached by others \citep{AAE,Pascale2018,Read2019}. However, this is sensitive to how one allocates membership of the dSph -- for instance, \cite{Lokas2009} and \cite{AmoriscoEvans2012} find negative excess kurtosis, possibly due to overly strict thresholds on membership, whilst \cite{Breddels2013} find the excess kurtosis is approximately zero.

\begin{figure}
    \centering
    \includegraphics[width=\columnwidth]{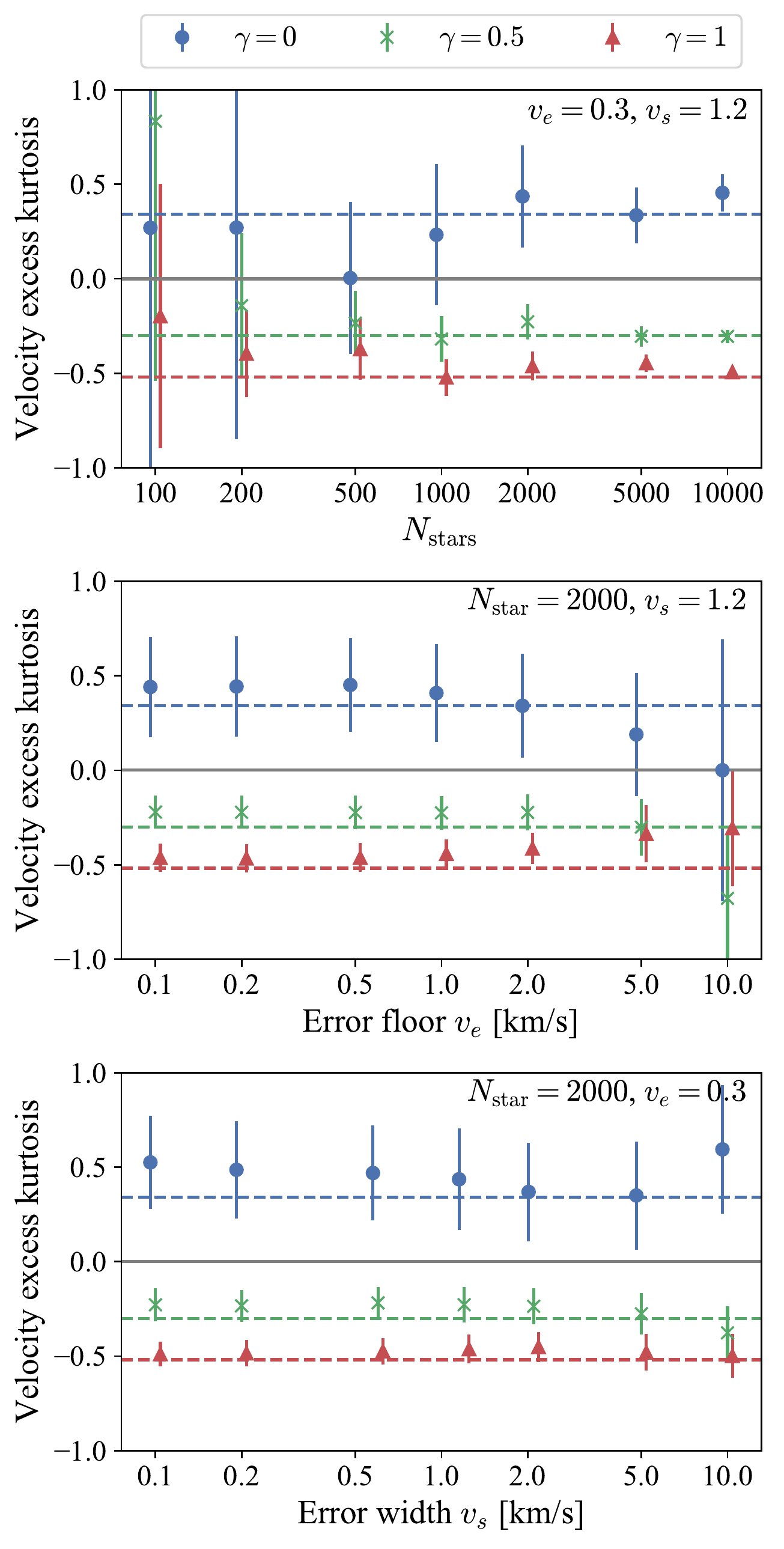}
    \caption{Measurement of velocity excess kurtosis using discrete samples of dynamical dwarf spheroidal models. Three models are inspected (cored = blue dot, weak cusp = green cross, strong cusp = red triangle). In the top panel the number of tracers is varied, central panel the line-of-sight velocity error floor is varied and bottom panel the width of the velocity error distribution varied. The dashed lines show the true values.}
    \label{fig:uncertainties}
\end{figure}
\begin{figure*}
    \centering
    \includegraphics[width=.97\textwidth]{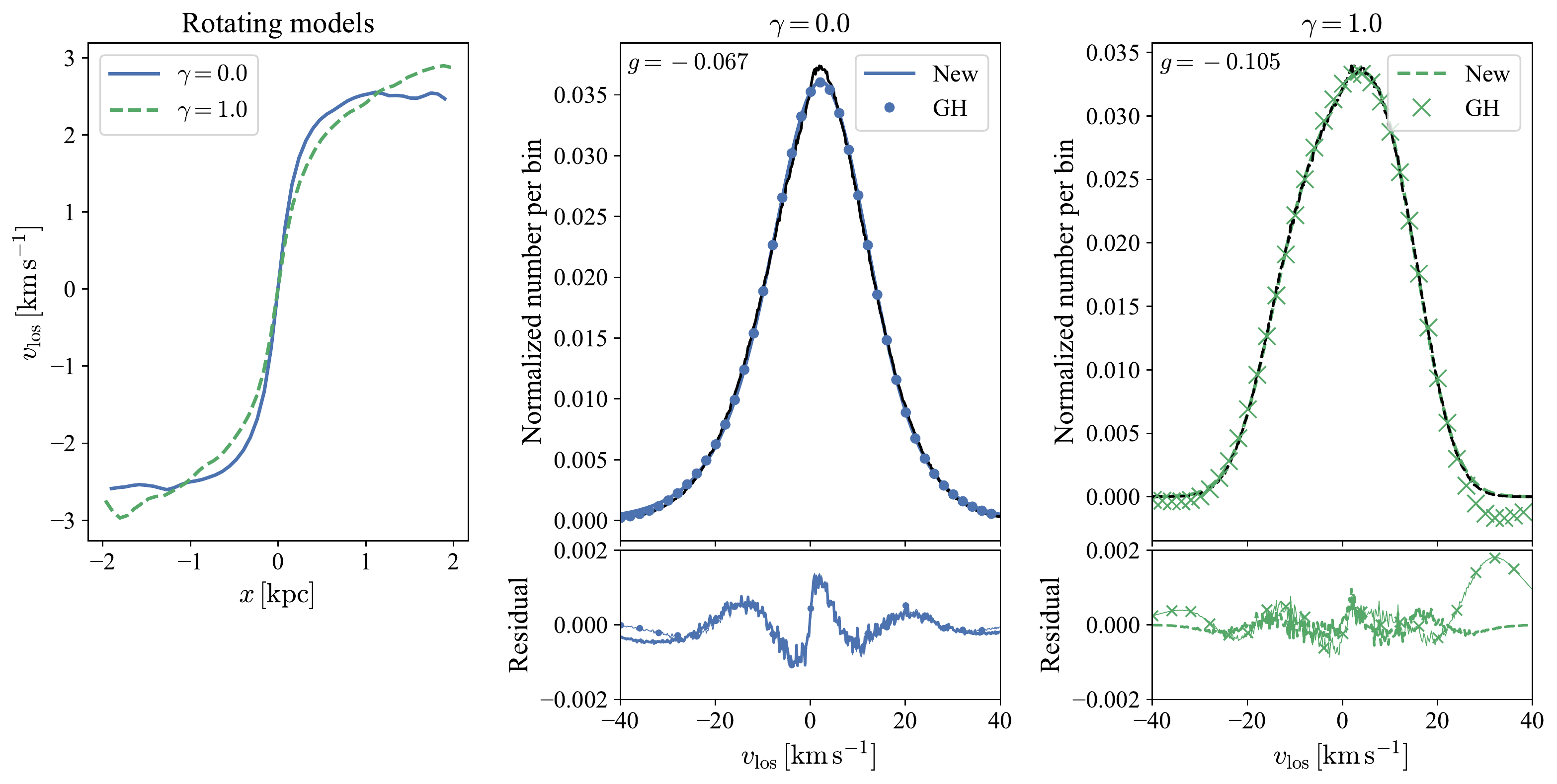}
    \caption{Example of rotating dynamical dwarf spheroidal models. Left panel shows the rotation curve of a cored (blue solid) and cuspy (green dashed) model. $x$ is the on-sky coordinate perpendicular to the rotation axis. The right two panels show the velocity distributions $\mathrm{sgn}(x)v_\mathrm{los}$ in black along with fits using the new models of this paper (lines) and the Gauss-Hermite series (with dots and crosses). The skewness $g$ is given in each panel.}
    \label{fig:rotating_models}
\end{figure*}

\subsection{Extracting kurtosis profiles}\label{sec::kurtfit}

With our dynamical models for Fornax in place, we now wish to extract the kinematic properties as a function of location in the dwarf spheroidal given known uncertainties. For this we use the newly introduced models. For the data, we avoid applying arbitrary cuts to the data, and instead employ a mixture model for the dwarf plus background Milky Way contaminants. To incorporate more information on dwarf membership, we use the Mg triplet equivalent width measurements from \cite{Walker2009} in addition to the velocities. When these measurements are unavailable we assign a default value with an uncertainty of $1000\,$\AA. 

\subsubsection{Chemo-kinematic mixture model}

The likelihood for an individual star at (circularized) radius $R$ with velocity $(v\pm\sigma_v)$ and equivalent width $(W\pm\sigma_W)$ is
\begin{equation}
\begin{split}
    \mathcal{L}&(v,W|\sigma_v,\sigma_W,R) =\\ &p_\mathrm{mem}(R)f_{\sigma_v}(v|\mu(R),\sigma(R),h_4(R))\mathcal{N}(W-W_\mathrm{d}|\sqrt{\sigma_W^2+\sigma_{W\mathrm{d}}^2}) \\&+ (1-p_\mathrm{mem}(R))B(v|\sigma_v)\mathcal{N}(W-W_\mathrm{o}|\sqrt{\sigma_W^2+\sigma_{W\mathrm{o}}^2}),
\end{split}
\label{eqn:ck_like}
\end{equation}
where $f_{\sigma_v}(v|\mu(R),\sigma(R),h_4(R))$ are the models introduced in this paper and the functions $\mu(R)$, $\sigma(R)$ and $h_4(R)$ are the parameters of the line-of-sight velocity distribution as a function of radius (mean, dispersion parameter and $h_4$). We assume $h_3=0$ everywhere. We use interpolating splines for $\mu(R)$, $\sigma(R)$ and $h_4(R)$ specifying $\mu_k$, $\sigma_k$ and $h_{4k}$ at 4 radii equally spaced in radius percentiles between the $1$st and $99$th of the data. 

For the velocity distribution of the background Milky Way contaminants $B(v|\sigma_v)$, we use Galaxia~\citep{Sharma2010} to sample stars in a $8\,\mathrm{deg}^2$ field around Fornax and in the magnitude range $18<V<20$. To convert the samples into a probability density function, we utilise a Gaussian kernel density estimate with a minimum kernel size of $3\,\mathrm{km\,s}^{-1}$. For a range of uncertainties $\sigma_v$, we use a broader kernel of $\sqrt{(3\,\mathrm{km\,s}^{-1})^2+\sigma_v^2}$ and compute the velocity and uncertainty dependent background model $B(v|\sigma_v)$ for each star. $\mathcal{N}(W-W_i|\sqrt{\sigma_W^2+\sigma_{Wi}^2})$ are Gaussians for the equivalent width measurements with free means $W_i$ and widths $\sigma_{Wi}$. The probability of membership $p_\mathrm{mem}$ is a function of on-sky location
\begin{equation}
p_\mathrm{mem}(R) = 1-\frac{n_\mathrm{b}}{n_\mathrm{b}+n_\mathrm{d}\Sigma_\mathrm{d}(R)/\Sigma_\mathrm{d}(100\,\mathrm{pc})},
\end{equation} 
where $n_\mathrm{b}$ is the on-sky background density from Galaxia ($0.23$ stars$\,$/$\,$arcmin$^2$), $n_\mathrm{d}$ is a free parameter giving the on-sky dwarf central density and $\Sigma_\mathrm{d}(R)$ is the median normalized density profile from \cite{Moskowitz2020}.

We adopt uniform priors on $\mu_k$, $\ln\sigma_k$ and $h_{4k}$, where we work with a tanh transformation of $h_4$ to ensure it stays within the required range $-0.188<h_4<0.145$. For the equivalent-width models, we employ uniform priors for $W_i$ and $\ln\sigma_{Wi}$. We use a uniform prior for the logarithm of the dwarf number density, $\ln n_\mathrm{d}$. We therefore have a total of $17$ parameters: $3$ groups of $4$ spline points $(\{\mu_k\},\{\ln\sigma_k\},\{h_{4k}\})$, $2$ sets of equivalent-width mean and dispersion $(\{W_i\},\{\ln\sigma_{Wi}\})$ and the density $n_\mathrm{d}$. 
We sample from the product of the likelihood in equation~\eqref{eqn:ck_like} over stars using the \textsc{emcee} package \citep{emcee}. We focus on the results for $\sigma(R)$ and $h_4(R)$ below. 

For the other parameters, we find the mean velocity of Fornax is $(55.3\pm0.2)\,\mathrm{km\,s}^{-1}$ (computed by inverse-variance weighting of the spline points), consistent with that reported by \cite{McConnachie2012}. We find an indication of radial variation of the mean velocity with the central regions ($\lesssim100\,\mathrm{pc}$) at a velocity $\sim2\,\mathrm{km\,s}^{-1}$ higher. However, all spline points have consistent mean velocities within their uncertainties. We re-ran our models with a mean which doesn't vary radially and find the conclusions on the variance and kurtosis profiles unchanged. As found by \cite{Walker2009b}, Fornax stars are not well separated from the background stars in the Mg equivalent width measurement. For Fornax, we find $W_\mathrm{d}=(0.616^{+0.006}_{-0.006})$\AA\,and $\sigma_{W\mathrm{d}}=(0.109^{+0.005}_{-0.005})$\AA, whilst for the background distribution we find
$W_\mathrm{o}=(0.52^{+0.12}_{-0.13})$\AA\,and $\sigma_{W\mathrm{o}}=(0.34^{+0.11}_{-0.07})$\AA. Finally, for the central density of Fornax we find $n_\mathrm{d}=(20.8^{+3.1}_{-2.5})$ stars$\,/\,\mathrm{arcmin}^2$. This number obviously depends on the magnitude selection of stars so it is more informative to compare relative to the background density, assumed to be $n_\mathrm{b}=0.23$ stars$\,$/$\,$arcmin$^2$. We find $n_\mathrm{d}/n_\mathrm{b}\approx90$, in line with the results from \cite{Battaglia2006} (who find $n_\mathrm{d}/n_\mathrm{b}\approx(72/0.78)=93$) or \cite{Wang2019} (who find $n_\mathrm{d}/n_\mathrm{b}\approx85$).

\subsubsection{Results}

The results of our procedure for the Fornax data are shown in the top panels of Fig.~\ref{fig:fornax_fit}. In agreement with the simple weighted binned estimates (black points), we find that the data favours positive excess kurtosis models. The uncertainties are larger than the binned estimates as they also incorporate uncertainty from the measurement error and membership probabilities. We have also run our fitting using the Gauss-Hermite series for $h_4>0$ (where it is positive-definite). The results are shown in Fig.~\ref{fig:fornax_GH}. The recovered $h_4$ profile is very similar to using the Laplace kernel (see Fig.~\ref{fig:fornax_gh_comparison}) but the excess kurtosis is smaller, partly because the Gauss-Hermite series does not admit excess kurtosis larger than $\sim0.5$.

To test our procedure, we also attempt to recover the kinematic profiles from our mock dynamical data. For this, we require an approximation to the uncertainty distribution, for which we use a gamma distribution $(\sigma_v-v_e)/v_s\sim\Gamma(1.1,1)$. $v_e$ is an error floor and $v_s$ governs the width of the tail of the error distribution. We set $v_e=0.3\,\mathrm{km\,s}^{-1}$ and $v_s=1.2\,\mathrm{km\,s}^{-1}$, which gives a good match to the error distribution of the Fornax sample. From our mock dynamical models, we select $N_\mathrm{stars}=2000$ stars, assign uncertainties and scatter the velocities by the uncertainty. We do not use a background model and set $p_\mathrm{mem}=1$ for all mock data. We show the results of applying our modelling to the three models in the lower panels of Fig.~\ref{fig:fornax_fit}. We find the recovery of the dispersion and excess kurtosis or $h_4$ is satisfactory.

We run further tests on the recovery of the excess kurtosis of our mock dynamical models with the quality and size of the dataset. For speed, we ignore the radial dependence of $\mu$, $\sigma$ and $h_4$. Additionally, we set a prior lower limit on $\sigma=0.05\,\mathrm{km\,s}^{-1}$. We run a set of three experiments varying (i) the number of stars $N_\mathrm{stars}$, (ii) the error floor, $v_e$, and (iii) the scale of the velocity error distribution, $v_s$. We use default parameters of $N_\mathrm{star}=2000$, $v_s=1.2\,\mathrm{km\,s}^{-1}$ and $v_e=0.3\,\mathrm{km\,s}^{-1}$ as used in the previous test. We use a fixed random seed to generate the mock samples from the models. The results of these experiments are shown in Fig.~\ref{fig:uncertainties}. We have varied $N_\mathrm{star}$ from $100$ to $10000$ and find that significant negative excess kurtosis can be found with relatively few stars ($\approx200$), but for a significant positive excess kurtosis measurement we require $\gtrsim2000$ stars. We have varied the error floor from $v_e=0.1$ up to $v_e=10\,\mathrm{km\,s}^{-1}$ finding that only for $v_e\lesssim2\,\mathrm{km\,s}^{-1}$ can the sign of the excess kurtosis be reliable determined, although again negative excess kurtosis is detectable for $v_e\sim5\,\mathrm{km\,s}^{-1}$. Finally, we have varied $v_s$ from $0.1$ to $10\,\mathrm{km\,s}^{-1}$ finding its effect on the recovery and size of the uncertainties to be quite weak such that for all of our tests we can fairly reliably distinguish the sign of the excess kurtosis and for $v_s\lesssim2\,\mathrm{km\,s}^{-1}$ the models are distinguishable. Although our results are tailored for Fornax, they can be extended to other dwarf spheroidals by scaling by the dispersion of Fornax of $11\,\mathrm{km\,s}^{-1}$.

\subsection{Extension to rotation}

To conclude our investigation into the use of the new models in the modelling of dwarf spheroidal data, we briefly demonstrate their application to rotating galaxies. Typically dwarf spheroidals exhibit negligible rotation and most detected rotation can be attributed to perspective effects from the projection of the bulk proper motion \citep{Walker2008,AmoriscoEvans2012}. However, if rotation is present, the combined signals of shift in the mean velocity and skewness may prove powerful for its separation from perspective rotation \citep{AmoriscoEvans2012}.

We modify the dynamical models by multiplying by an odd function of the $z$-component of angular momentum, $J_\phi$,
\begin{equation}
    f_\mathrm{rot}(\bs{J})=f(\bs{J})\Big(1 + k\,\mathrm{sgn}(J_\phi)\Big).
\end{equation}
$k$ controls the degree of rotation and can take values $|k|\leq1$. We investigate the cored $\gamma=0$ and cusped $\gamma=1$ models from the previous section. We choose $k=0.3$ for the cored model and $k=0.2$ for the cusped model, and view the models perpendicular to the rotation axis. 

In Fig.~\ref{fig:rotating_models}, we show the rotation curves of the two models. The rotation reaches $\sim3\,\mathrm{km\,s}$ for both models. Such a large rotation signal in Fornax would be obvious so this experiment is purely illustrative. We also display the line-of-sight velocity distributions (where the symmetry allows us to increase the number statistics by multiplying the velocities by the sign of $x$). We see both distributions are weakly skewed ($g=-0.067$ for the cored model and $g=-0.105$ for the cusped). Both the Gauss-Hermite series and the new models from this paper provide excellent fits to the distributions. In the case of the positive excess kurtosis model the fits are near identical. However, we see for the cusped negative excess kurtosis model the residuals with respect to our new models are in general smaller and in particular better capture the right-hand wing.

\section{Application to spectral fitting}\label{section::spectral}

\begin{figure*}
    \centering
    \includegraphics[width=.97\textwidth]{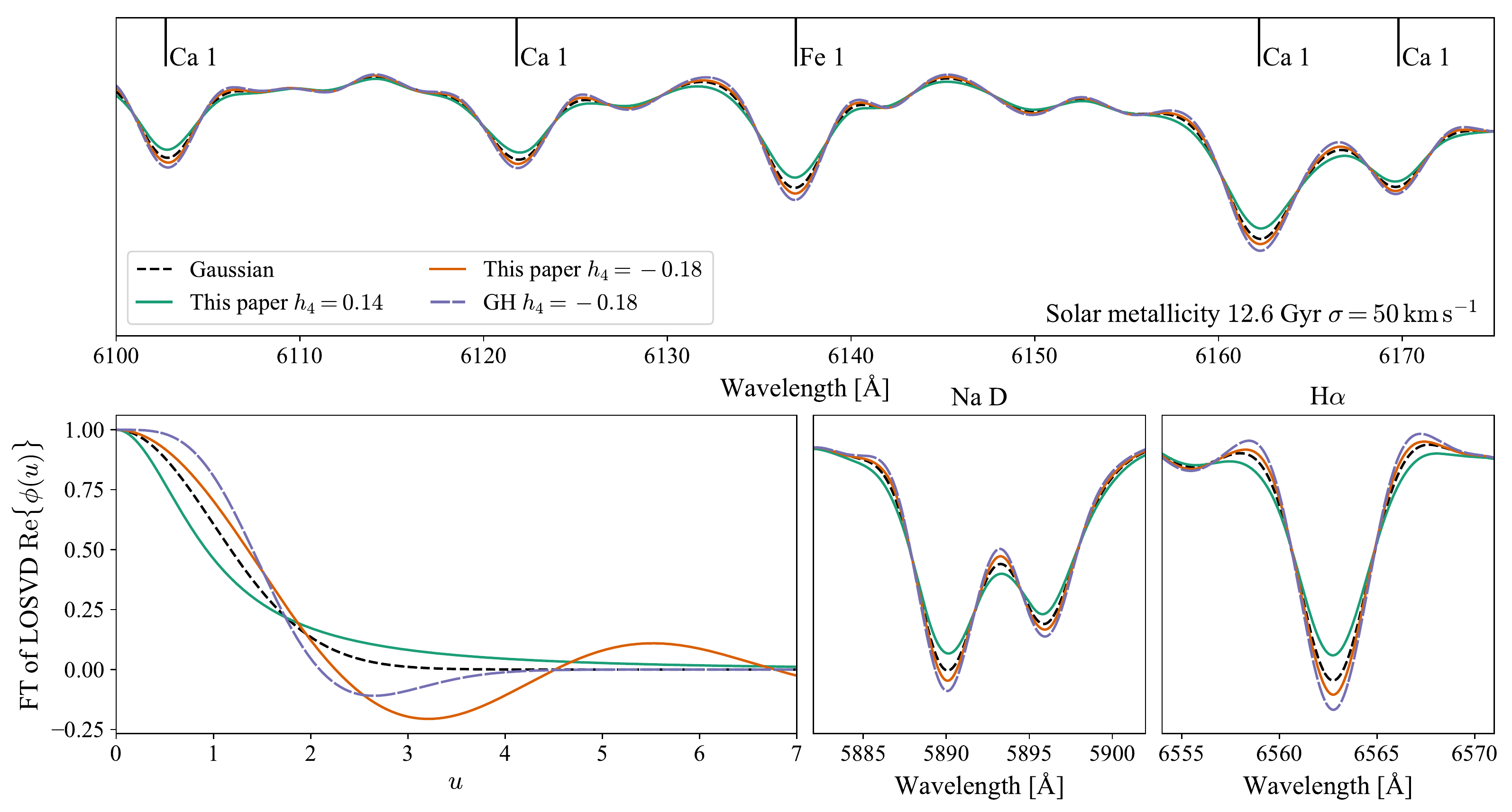}
    \caption{Single stellar population (solar metallicity, $12.6\,\mathrm{Gyr}$ convolved with different choices of line-of-sight velocity distribution (LOSVD). Four LOSVDs are used, all with identical $\sigma=50\,\mathrm{km\,s}^{-1}$ parameters (which does not equal the velocity dispersion except for the Gaussian): (i) Gaussian in black dashed, (ii) solid green this paper's model $h_4=0.14$, (iii) solid orange this paper's model $h_4=-0.18$ and (iv) Gauss-Hermite series with $h_4=-0.18$. The top panel and bottom right two panels show example regions of the spectra. The corresponding Fourier transforms of the LOSVDs (important in the computation of the spectra) are given in the bottom left. Note the selected $h_4$ parameters are extreme to exaggerate the effects.}
    \label{fig:spectra}
\end{figure*}

As discussed in Section~\ref{sec::GHSERIES}, the Gauss-Hermite series was originally introduced for the extraction of kinematics from spectra \citep{Gerhard1993,vanderMarelFranx1993}. Given a library of single stellar population spectra, \cite{CappellariEmsellem2004} describe a penalized likelihood algorithm for the simultaneous extraction of the age-metallicity properties of a measured galaxy and the kinematics via the convolution of the library with a Gauss-Hermite series for the LOSVD. \cite{Cappellari2017} showed how undersampling the LOSVD when the LOSVD dispersion is smaller than half the pixel spacing is problematic for accurate recovery of the kinematics (particularly the mean) but could be circumvented using the analytic Fourier transform of the Gauss-Hermite series. This is now the standard method in the spectral fitting code \textsc{ppxf} \citep{CappellariEmsellem2004,Cappellari2017}.

Although in the spectral fitting application the issue of negative wings in the LOSVD is less severe than in the application to discrete data, it still gives rise to unphysical features that could bias the recovery of kinematics. However, the new families of models introduced in this paper provide simple replacements for the Gauss-Hermite series in the spectral fitting procedure. This is because they too have analytic Fourier transforms. In fact, the Fourier transforms are simpler to evaluate than the real-space forms. The Fourier transforms for the two models are
\begin{equation}
    \phi_s(u) = \frac{1}{2}e^{iw_0u-u^2t^2/2b^2}\Big[\frac{1}{1-ia_+u/b}+\frac{1}{1+ia_-u/b}\Big],
\end{equation}
for the Laplace kernel, and
\begin{equation}
    \phi_s(u) = \frac{1}{2}e^{iw_0u-u^2t^2/2b^2}\Big[\frac{e^{ia_+u/b}}{ia_+u/b}-\frac{e^{ia_-u/b}}{ia_-u/b}\Big],
\end{equation}
for the uniform kernel. We have implemented the models within a version of the \textsc{ppxf} code and find that for the included examples the recovery of the kinematics is very similar to using a Gauss-Hermite series.

In Fig.~\ref{fig:spectra}, we show examples of a single stellar population with solar metallicity and of age $12.6\,\mathrm{Gyr}$ \citep{Vazdekis2010} convolved with different choices of LOSVD. For all models, we fix the parameter $\sigma=50\,\mathrm{km\,s}^{-1}$ but this does not mean the variance of all models are the same. We see that for $h_4>0$ the wings of the lines are broader than Gaussian and for $h_4<0$ they are narrower. We observe that when using the Gauss-Hermite series there are prominent bumps in the wings of the lines (particularly visible for H$\alpha$) due to the negative LOSVD but these are avoided when using the models of this paper. It should be noted however that in general the effects and differences are small and the large values of $|h_4|$ chosen for the example have been selected to exaggerate the effect. We also display the Fourier transform of the LOSVD. One feature of the Gauss-Hermite series is that it decays like $e^{-u^2}$ in Fourier space whereas our negative models oscillate and decay like $1/u$. However, this is unimportant as the template spectra are bandwidth-limited and Nyquist-sampled so we only require the Fourier transform on a fixed grid ($u/\sigma<\pi$) and its behaviour for large $u$ is irrelevant. The only disadvantages to our suggested modifications are: (i) one cannot incorporate higher order moments $h_5$ or $h_6$ -- these are already non-zero for our models. In practice, one rarely uses these higher order moments anyway; (ii) there is only a finite range of $h_3$ and $h_4$ accessible by our models. Already $h_3$ and $h_4$ are bounded in \textsc{ppxf} (default is between $-0.3$ and $0.3$) but the difficulty with our models is in the inter-dependence of maximum $h_3$ and $h_4$ as shown in Fig.~\ref{fig:parameter_space}. One could use $a$ and $\delta$ as parameters instead.

\section{Conclusions}

We have introduced a new method for constructing probability distribution functions for the modelling of discrete stellar velocity data drawn from weakly non-Gaussian distributions and with heteroskedastic uncertainties. Our study was motivated by the limitation of the commonly used Gauss-Hermite series expansion. Although the Gauss-Hermite series has attractive properties for convolution with measurement uncertainty, a key limitation is that the Gauss-Hermite series does not give well-defined probability density functions -- there are regions of negativity particularly for negative excess kurtosis models.

We have demonstrated a generic method for constructing everywhere positive families of probability functions through the convolution of a Gaussian with a choice of kernel. The resulting probability distribution function inherits the properties (moments and cumulants) of the kernel. We have examined in detail two choices of kernel, Laplace and uniform, which give rise to positive and negative excess kurtosis respectively and can be asymmetrised to incorporate skewness. Inspired by the standard implementation of the Gauss-Hermite expansion, we have restricted the models to have approximately $h_1=h_2=0$ such that the models can be used as a direct replacement for the Gauss-Hermite series. However, in theory, the models can be used in greater generality. We provide code for the evaluation of our models at \href{https://github.com/jls713/gh_alternative}{https://github.com/jls713/gh\_alternative}. 

We have demonstrated the use of our new models with two simple applications. The first is to discrete sets of line-of-sight velocity data in galactic modelling. We have shown through a series of dynamical models how non-Gaussianity can break mass-anisotropy degeneracies and how our methods allow for a flexible measurement of the excess kurtosis. This has a ready application to the dwarf galaxies around the Milky Way, for which large datasets of velocities of bright giant stars exist and for which the structure of the dark halo is an open question. For the specific example of the Fornax dwarf spheroidal galaxy, we have shown using a chemo-kinematical mixture model that the data seem to favour positive excess kurtosis velocity distributions and hence indicate cored dark matter density profiles. Additionally, we have tested our modelling procedure on a series of dynamical dwarf spheroidal galaxy models with varying dark matter profiles, finding good recovery of the kinematic profiles and that a cored model well reproduces the 
observed data. Furthermore, we have investigated how the excess kurtosis measurement depends on the number and uncertainty distribution of stellar tracers. For a Fornax-like dSph. $\gtrsim2000$ stars with uncertainties $\lesssim2\,\mathrm{km\,s}^{-1}$ are sufficient to measure the sign of the excess kurtosis and measure the mass density slope, although this assumes perfect knowledge of dwarf spheroidal membership. We have also shown how our models are able to capture skewness, which could be a powerful indicator of the subtle effects of rotation and immune to the effects of perspective.

Secondly, although our models were not motivated by spectral fitting applications, their analytic Fourier transforms allow them to be used in spectral fitting algorithms where they could reduce bias produced by negative wings in the line-of-sight velocity distributions of high quality data.

Our methods for characterising mildly non-Gaussian velocity distributions are flexible and adaptable, offering significant advantages over the conventional Gauss-Hermite series. The skewness and kurtosis of such distributions contain important clues on the orbits of the tracers and the gravitational potential in which they move. The sizes of such discrete velocity datasets -- whether for populations in the Milky Way or its dwarf satellite entourage or for populations in external galaxies like globular clusters and planetary nebulae -- have substantially increased in recent years. We envisage that our new methods will play an important role in exploiting these datasets to the full.

\section*{Acknowledgements}
JLS acknowledges support from the Royal Society (URF\textbackslash R1\textbackslash191555). 
This paper made use of
\textsc{Mathematica} \citep{mathematica}, 
\textsc{numpy} \citep{numpy},
\textsc{scipy} \citep{scipy}, 
\textsc{matplotlib} \citep{matplotlib}, 
\textsc{seaborn} \citep{seaborn},
\textsc{astropy} \citep{astropy:2013,astropy:2018},
\textsc{emcee} \citep{emcee},
\textsc{agama} \citep{Vasiliev2019} and \textsc{ppxf} \citep{Cappellari2017}. 
This work has made use of data from the European Space Agency (ESA) mission
{\it Gaia} (\url{https://www.cosmos.esa.int/gaia}) \citep{Gaia2016,Gaia2018}, processed by the {\it Gaia}
Data Processing and Analysis Consortium (DPAC,
\url{https://www.cosmos.esa.int/web/gaia/dpac/consortium}). Funding for the DPAC
has been provided by national institutions, in particular the institutions
participating in the {\it Gaia} Multilateral Agreement.

\section*{Data availability}
All data used in this article are in the public domain: 
\cite{Walker2009}, 
\cite{Vazdekis2010},
\cite{HelmiGaiaDR2},
\cite{Wang2019} and
\cite{Moskowitz2020}.
%%%%%%%%%%%%%%%%%%%%%%%%%%%%%%%%%%%%%%%%%%%%%%%%%%

%%%%%%%%%%%%%%%%%%%% REFERENCES %%%%%%%%%%%%%%%%%%

\bibliographystyle{mnras}
\bibliography{bibliography} 
%%%%%%%%%%%%%%%%%%%%%%%%%%%%%%%%%%%%%%%%%%%%%%%%%%

%%%%%%%%%%%%%%%%% APPENDICES %%%%%%%%%%%%%%%%%%%%%

\appendix
\section{Hermite polynomials}\label{appendix::hermite}
The convention for the Hermite polynomials employed here follows \cite{vanderMarelFranx1993} such that they are given by the relation
\begin{equation}
    \Big(-\frac{\mathrm{d}}{\mathrm{d}y}\Big)^l\alpha(y) = \sqrt{l!}H_l(y/\sqrt{2})\alpha(y).
\end{equation}
This is different from the standard $H_n(y)$ defined in e.g. \cite{Olver2010} which are given by $\sqrt{l!2^l}H_l(y)$. The first five Hermite polynomials are
\begin{equation}
    \begin{split}
        H_0(y)&=1,\>\>
        H_1(y)=\sqrt{2}y,\>\>
        H_2(y)=\frac{1}{\sqrt{2}}(2y^2-1),\\
        H_3(y)&=\frac{1}{\sqrt{6}}(2\sqrt{2}y^3-3\sqrt{2}y),\>\>
        H_4(y)=\frac{1}{\sqrt{24}}(4y^4-12y^2+3).\\
    \end{split}
\end{equation}
The generating function for this definition of the Hermite polynomials is given by
\begin{equation}
    \sum_{n=0}^{n=\infty}\frac{z^n}{n!}\sqrt{n!}H_n(y) = e^{\sqrt{2}yz-z^2/2},
    \label{eqn::hp_gf}
\end{equation}
from which the $n$th polynomial can be found as
\begin{equation}
    H_n(y) = \frac{1}{\sqrt{n!}}\frac{\partial^n}{\partial z^n}\Big(e^{\sqrt{2}yz-z^2/2}\Big)\Big|_{z=0}.
\end{equation}

\section{Cumulants of the Gauss-Hermite series}\label{appendix::gh_cumulants}
The cumulant generating function of the Gauss-Hermite series given in equation~\eqref{eqn::gauss_hermite} is given by
\begin{equation}
\begin{split}
    \ln\phi_\mathrm{GH}(u) &= \ln\int_{-\infty}^{\infty}\mathrm{d}x\,e^{iux}f(x) = \sum_{r=1}\kappa_{r}\frac{(iu)^r}{r!},
    \\&=iVu-\frac{(\sigma u)^2}{2}+\ln\Big[1+\sum_{n\geq3}i^nh_nH_n(\sigma u)\Big].
\end{split}
\end{equation}
The cumulants are then given by
\begin{equation}
\begin{split}
    \kappa_1&=V+\tilde\kappa_1,\>\>
    \kappa_2=\sigma^2+\tilde\kappa_2,\>\>
    \kappa_i=\tilde\kappa_i,\quad i\geq3,\\
    \tilde\kappa_r &= \sigma^r\frac{\mathrm{d}^{r-1}}{\mathrm{d}u^{r-1}}\Big[\Big(1+\sum_{n\geq3}i^nh_nH_n( u)\Big)^{-1}\sum_{n\geq3}i^{n-r}h_n\frac{\mathrm{d}H_n(u)}{\mathrm{d}u}\Big]_{u=0}.
\end{split}
\end{equation}
Performing the derivative produces long-winded expressions. For their evaluation, we require the Hermite numbers $H_n(0)$ given by
\begin{equation}
    H_n(0) = (-1)^{n/2}\frac{\sqrt{n!}}{(n/2)!\sqrt{2^n}},
\end{equation}
for even $n$ and zero for odd $n$, and the $n$th derivative of the Hermite polynomials is
\begin{equation}
    \frac{\mathrm{d}^rH_n(x)}{\mathrm{d}x^r}\Big|_{x=0} = \sqrt{\frac{2^rn!}{(n-r)!}}H_{n-r}(0)=(-1)^{(n-r)/2}\frac{2^{r-n/2}\sqrt{n!}}{((n-r)/2)!},
\end{equation}
for even $n-r$ and zero otherwise. Restricting to $h_n=0$ for $n>4$, we find for the first four cumulants (note all $\kappa_r$ are non-zero)
\begin{equation}
    \begin{split}
        \kappa_1 &= V + \sqrt{3}\sigma\lambda h_3\approx V + \sqrt{3}\sigma h_3,\\
        \kappa_2 &= \sigma^2 +\sigma^2\lambda^2(h_4(2\sqrt{6}+3h_4)-3h_3^2)\approx\sigma^2(1+2\sqrt{6}h_4),\\
        \kappa_3 &= \tfrac{1}{2}h_3\sigma^3\lambda^3\Big(8\sqrt{3}+12\sqrt{3}h_3^2-3h_4(8\sqrt{2}+5\sqrt{3}h_4)\Big)\approx4\sqrt{3}h_3\sigma^3,\\
        \kappa_4 &= \tfrac{1}{2}\sigma^4\lambda^4\Big(16\sqrt{6}h_4-9h_4^2(8+6\sqrt{6}h_4+5h_4^2)\\&+12h_3^2(15h_4^2+8\sqrt{6}h_4-8)-108h_3^4\Big)\approx8\sqrt{6}h_4\sigma^4,
    \end{split}
\end{equation}
where $\lambda = \Big(1+\sqrt{3/8}h_4\Big)^{-1}$ and the approximations for small $h_n$ agree with \cite{vanderMarelFranx1993}. Note that for larger values $|h_4|$, the excess kurtosis $\kappa=\kappa_4/\kappa_2^2$ is not a monotonic function of $h_4$.

\section{Convolution of a Gauss-Hermite series with Gaussian errors}\label{appendix::convolution}
In this appendix, we compute the convolution of a Gauss-Hermite series $f(x)$ with a Gaussian uncertainty distribution $\mathcal{N}(x-x'|\sigma_e)$ as
\begin{equation}
    f_{\sigma_e}(x)=\int_{-\infty}^{\infty}\mathrm{d}x'\,f(x')
    \mathcal{N}(x-x'|\sigma_e).
    \label{eqn::convGH}
\end{equation}
We work in terms of the scaled and shifted coordinate $w=(x-V)/\sigma$. Dropping the constant factors, we require the convolution
\begin{equation}
    f_s(w) = \int_{-\infty}^\infty\mathrm{d}w'\,f(w')\mathcal{N}(w-w'|s),
\end{equation}
where
\begin{equation}
    \mathcal{N}(w|s)=\frac{1}{\sqrt{2\pi s^2}}e^{-\tfrac{1}{2}w^2/s^2},
\end{equation}
and $s=\sigma_e/\sigma$. For this computation, we shift to Fourier space. One attractive property of the Gauss-Hermite series is that the terms are eigenfunctions of the Fourier transform, $F$,
\begin{equation}
    F\Big\{\alpha(w)H_n(w)\Big\} \equiv \int_{-\infty}^\infty\mathrm{d}w\,\alpha(w)H_n(w)e^{ikw} = \sqrt{2\pi}i^nH_n(k)\alpha(k).
\end{equation}
\cite{Cappellari2017} uses this property to rapidly convolve a model spectrum with the Gauss-Hermite series. Each term in the convolution of $\mathcal{N}(w|s)$ with $f(w)$ is given by
\begin{equation}
    \mathcal{C}_n(w|s)\equiv\int\mathrm{d}w'H_n(w')\alpha(w')\mathcal{N}(w-w'|s).
    \label{eqn::cn_fourier_transform}
\end{equation}
Taking the Fourier transform of the convolution yields
\begin{equation}
    F\Big\{\mathcal{C}_n(w|s)\Big\}
    =\sqrt{2\pi}i^nH_n(k)\alpha(k)e^{-s^2k^2/2}
    = \sqrt{2\pi}i^nH_n(k)\alpha(s'k),
\end{equation}
where $s'=\sqrt{1+s^2}$. The inverse Fourier transform is given by
\begin{equation}
    \mathcal{C}_n(w|s)
    =\frac{i^n}{\sqrt{2\pi}s'}\int\mathrm{d}k'e^{-ik'\frac{w}{s'}}H_n\Big(\frac{k'}{s'}\Big)\alpha(k').
\end{equation}
The Hermite polynomials satisfy the following multiplication theorem \citep[Eq.~18.18.13, ][]{Olver2010}
\begin{equation}
    H _ { n } ( \gamma x ) = \gamma ^ { n } \sum _ { j = 0 } ^ { \lfloor \frac { n } { 2 } \rfloor } ( 1 - \gamma ^ { -2 } ) ^ { j } \frac{\sqrt{n!}}{j!2^{j}\sqrt{(n-2j)!}} H _ { n - 2 j } ( x ).
\end{equation}
Rewriting $H(k'/s')$ using this series and using the eigenfunction of the Fourier transform property of $H(k')\alpha(k')$ gives an expression for the convolution of each term as
\begin{equation}
    \mathcal{C}_n(w|s) = \frac{\mathcal{N}(w|s')}{s'^n}\sum _ { j = 0 } ^ { \lfloor \frac { n } { 2 } \rfloor }s^{2j} \frac{\sqrt{n!}}{j!2^{j}\sqrt{(n-2j)!}}H _ { n - 2 j } \Big(\frac{w}{s'}\Big).
    \label{eqn::cn}
\end{equation}
The sum can be recombined into a single Hermite polynomial as
\begin{equation}
    \mathcal{C}_n(w|s) = \frac{\mathcal{N}(w|s')}{s'^n}(1-s^2)^{n/2}H_n\Big(\frac{w}{\sqrt{1-s^4}}\Big).
    \label{eqn::cn_simpler}
\end{equation}
This formula can be derived directly from equation~\eqref{eqn::cn_fourier_transform} using equation (7.374.8) of \cite{gradshteyn2007}. Note that the formula is valid for $s>1$ although it requires numerical care, as both the prefactor and the argument of the Hermite polynomial are imaginary in this case.
We give the full result for $f_{\sigma_e}(x)$ in equation~\eqref{eqn::convGH_fullresult}.

\section{Other series methods}\label{appendix::other_series}
The Gauss-Hermite series bears resemblance to two other series expansions that are more commonly used in probability theory. The Gram-Charlier and Edgeworth series both approximate a general probability density function in terms of its cumulants \citep[see][for general derivations]{Blinnikov1998}. The construction of the two series are similar in that the cumulant generating function is written as a sum of a Gaussian cumulant generating function (first two terms) plus a truncated cumulant generating function series (for $n\geq3$):
\begin{equation}
    \ln\phi_\mathrm{approx}(u)=\ln\phi_\mathrm{Gaussian}(u)+\sum_{r\geq3}\kappa_r\frac{(iu)^r}{r!}.
\label{eqn::series_characterisitic_function}
\end{equation}
The approximating probability density function is then the inverse Fourier transform of $\phi_\mathrm{approx}$. The difference between Gram-Charlier and Edgeworth is in how one chooses to truncate the sum. In Gram-Charlier, the sum is truncated at the $n$th cumulant, whilst for Edgeworth the expansion of the normalized variable $x/\sigma$ is truncated at the $n$th power of the standard deviation, $\sigma$ (making it a truly asymptotic series). The advantage of this approach is the terms in the resulting series are simply related to the cumulants (see Appendix~\ref{appendix::gh_cumulants} for the relationship between the Gauss-Hermite coefficients and the cumulants). 
Up to fourth order, the Gram-Charlier series is given by
\begin{equation}
f(x) = \frac{1}{\sigma}\alpha(w)\Big[1+\frac{\kappa_3}{3!\sigma^3}He_3(w)+\frac{\kappa_4}{4!\sigma^4}He_4(w)\Big].
\label{eqn::GC}
\end{equation}
Correspondingly, the Edgeworth series is given by
\begin{equation}
f(x) = \frac{1}{\sigma}\alpha(w)\Big[1+\frac{\kappa_3}{3!\sigma^3}He_3(w)+\frac{\kappa_4}{4!\sigma^4}He_4(w)+\frac{10\kappa_3^2}{6!\sigma^6}He_6(w)\Big].
\label{eqn::Edgeworth}
\end{equation}
Here the Chebyshev-Hermite polynomials $He_n(y)$ are given by
\begin{equation}
    \begin{split}
        He_3(y)&=y^3-3y,\>\>
        He_4(y)=y^4-6y^2+3,\\
        He_6(y)&=y^6-15y^4+45y^2-15.\\
    \end{split}
\end{equation}

For both series, the convolution with observational uncertainties, $\sigma_e$, is simple as the series are derived directly from a characteristic function. The convolution is a multiplication in Fourier space such that $\phi_\mathrm{Gaussian}(u)$ in equation~\eqref{eqn::series_characterisitic_function} has a width $\sigma'=\sqrt{\sigma^2+\sigma_e^2}$. Equations~\eqref{eqn::GC} and~\eqref{eqn::Edgeworth} are then modified as $w\rightarrow w(\sigma/\sigma')$ and $\sigma\rightarrow\sigma'$.

The Gram-Charlier series has very poor convergence properties ($p(x)$ must fall off faster than $e^{-x^2/4}$ for large $x$) whilst the Edgeworth expansion is designed to converge given any $p(x)$. On the other hand, the Gauss-Hermite series will converge if $p(x)$ satisfies an $\alpha$-H\"older condition and $\int\mathrm{d}x\,|p(x)|(1+|x|^{-3/2})$ doesn't diverge \citep{Blinnikov1998}.

We observe that both series methods also necessarily suffer negative probability density regions so are not suitable work-arounds for our application. 

\section{Other choices of kernel}
In this appendix we give two tables detailing possible other choices of kernel. Table~\ref{table::kernels} gives the set of half-kernels, $K_+(y)$, which are non-zero for positive $y$ and their corresponding error-convolved distributions $f_{s+}(w)$, and characteristic function $\phi_{K+}(u)$. Table~\ref{table::cumulants} gives the cumulants for models formed by stitching together two of these half-kernels for positive and negative $y$ with different scale parameters.
\begin{table*}
\caption{Choices of half-kernel $K_+(y)$ (zero for $y<0$ -- all normalized to integrate to $\tfrac{1}{2}$ except Gamma which is normalized to unity), their error-convolved distribution $f_{s+}(w)$ and their characteristic functions $\phi_{K+}(u)$.
Models above the double line have negative excess kurtosis and positive below. 
$w'=w-w_0$, $t=1+b^2s^2$, $\mathcal{N}(w|s)$ is a normal distribution with width $s$ and $\Phi(x)$ is the cumulative distribution of the unit normal. ${}_1F_1(a,b,c,x)$ is the confluent hypergeometric function.
}
\begin{threeparttable}
\begin{tabular}{l|lll}
         \hline\\
     Name&$K_+(y)$&$f_{s+}(w)$&$\phi_{K+}(u)$  \\
     \hline\\
     Uniform&
$
\frac{1}{2a},\quad\mathrm{if}\,y<a,
$
&
$
\!\begin{aligned}[t]
\frac{b}{2a}\Big[\Phi\Big(\frac{bw'}{t}\Big)-\Phi\Big(\frac{bw'-a}{t}\Big)\Big]
\end{aligned}
$
&
$
\frac{e^{iau}-1}{2iau}
$
\\\\
\hline\\
     Cosine&
$
\frac{\pi}{4a}\cos\Big(\frac{\pi y}{2a}\Big),\quad\mathrm{if}\,y<a,
$
&
$
\frac{b\pi}{8a}\Big(D(w',w'-a/b)-D(-w',-w'+a/b)\Big)
$\footnotemark
&
$
\frac{\pi}{2}\frac{\pi e^{iau}-2iau}{\pi^2-4a^2u^2}
$\\\\
\hhline{====}
\\
     Laplace&
$
\frac{1}{2a}e^{-y/a}
$
&
$
\frac{b}{4a}\exp\Big(\frac{t^2-2abw'}{2a^2}\Big)\mathrm{erfc}\Big(\frac{t^2-abw'}{\sqrt{2}ta}\Big)
$
&
$
\frac{1}{2-2iau}
$
\\
\hline\\
Gaussian&
$
\frac{1}{\sqrt{2\pi}a}e^{-y^2/2a^2}
$&
$
b\mathcal{N}(bw'|\sqrt{t^2+a^2})\Phi\Big(\frac{bw'a}{t\sqrt{t^2+a^2}}\Big)
$
&
$
e^{-a^2u^2/2}\Phi(iau)
$
\\\\
\hline\\
Gamma&
$
\frac{y^{\beta-1} e^{-y/a}}{a^\beta\Gamma(\beta)}
$
&
$
\!\begin{aligned}[t]
&\frac{\sqrt{2^{\beta-3}}b}{\sqrt{\pi} a^{\beta+1}\Gamma(\beta)}t^{\beta-2}e^{-b^2w^2/2t^2}\times\\&\Big[at\Gamma\Big(\frac{\beta}{2}\Big){}_1F_1\Big(\frac{\beta}{2},\frac{1}{2},\frac{(t^2-abw')^2}{2a^2t^2}\Big)\\&+\sqrt{2}{(abw'-t^2)}\Gamma\Big(\frac{1+\beta}{2}\Big){}_1F_1\Big(\frac{1+\beta}{2},\frac{3}{2},\frac{(t^2-abw')^2}{2a^2t^2}\Big)\Big]
\end{aligned}
$
&
$
\frac{1}{(1-iau)^\beta}
$\\\\
\hline
\end{tabular}
    \begin{tablenotes}
      \footnotesize
\item\footnotemark[1]
$
D(w,q)=e^{-\pi^2t^2/8a^2}e^{i\pi bw/2a}\Big[\Phi\Big(\frac{i\pi t^2+2abw}{2at}\Big)-\Phi\Big(\frac{i\pi t^2 + 2abq}{2at}\Big)\Big]
$
\end{tablenotes}
\end{threeparttable}
\label{table::kernels}
\end{table*}

\begin{table*}
\caption{Cumulants for kernels built from two half-kernels from Table~\ref{table::kernels} with the requirement $\int_0^\infty\mathrm{d}y\,K_+(y)=\int_{-\infty}^0\mathrm{d}y\,K_-(y)=\tfrac{1}{2}$ (except Gamma where we only consider positive $y$). The width of the positive domain is $a_+$ and $a_-$ for the negative domain. These quantities can be derived from Table~\ref{table::kernels} using $\tilde\kappa_r=i^{-r}\partial^r_u\ln\Big[\phi_{K+}(u)+\phi_{K-}(u)\Big]_{u=0}$. We define the auxiliary variables. $a=\tfrac{1}{2}(a_++a_-)$ and $\Delta=\tfrac{1}{2}(a_+-a_-)$. $\tilde\mu$ is the mean, $\tilde v$ the variance and $\tilde\kappa_3$ and $\tilde\kappa_4$ the third and fourth cumulant from which the skewness $\tilde g$ and excess kurtosis $\tilde\kappa$ are computed as $\tilde g=\tilde\kappa_3/\tilde v^{3/2}$ and $\tilde\kappa=\tilde\kappa_4/\tilde v^2$. Note that all bracketed constants are positive.}
\begin{tabular}{l|llll|}
         \hline\\
     Name&$\tilde\mu$&$\tilde v$&$\tilde\kappa_3$&$\tilde\kappa_4$  \\
     \hline\\
     Uniform
      &$\frac{\Delta}{2}$
&$\frac{a^2}{3}+\frac{\Delta^2}{12}$
&$\frac{\Delta a^2}{4}$
&$-\frac{1}{120}\Big(16a^4-4a^2\Delta^2+\Delta^4\Big)$\\\\

Cosine
&$\Delta\Big(1-\frac{2}{\pi}\Big)$
&$
\Big(1-\frac{8}{\pi^2}\Big)a^2+\frac{4}{\pi}\Big(1-\frac{3}{\pi}\Big)\Delta^2
$
&$
\frac{2\Delta}{\pi}\Big[3\Big(\frac{4}{\pi}-1\Big)^2a^2+\Big(\frac{12}{\pi}-\frac{8}{\pi^2}-3\Big)\Delta^2\Big]
$
&
$
\!\begin{aligned}[t]
&-\frac{2a^4}{\pi^4}\Big(\pi^4-96\Big)+\frac{24a^2\Delta^2}{\pi^4}\Big(112-32\pi+2\pi^2-\pi^3\Big)\\&-\frac{8\Delta^4}{\pi^4}
\Big(12\pi^2-\pi^3-24\pi-12\Big)
\end{aligned}
$
\\\\
\hhline{=====}
\\
Laplace
&$\Delta$
&$2a^2+\Delta^2$
&$2\Delta(6a^2+\Delta^2)$
&$6\Big(2a^4+12a^2\Delta^2+\Delta^4\Big)$\\\\
Gaussian
&
$\sqrt{\frac{2}{\pi}}\Delta$
&
$
\!\begin{aligned}[t]
&a^2+\Big(1-\frac{2}{\pi}\Big)\Delta^2
\end{aligned}
$
&
$
\sqrt{\frac{2}{\pi}}\Delta\Big[3a^2+\Big(\frac{4}{\pi}-1\Big)\Delta^2\Big]
$
&
$
4\Delta^2\Big[3\Big(1-\frac{2}{\pi}\Big)a^2+\frac{2}{\pi}\Big(1-\frac{3}{\pi}\Big)\Delta^2\Big]
$
\\\\
Gamma&
$
a\beta
$
&
$
a^2\beta
$
&
$
2a^3\beta
$
&
$
6a^4\beta
$\\\\
\hline
\end{tabular}
\label{table::cumulants}
\end{table*}

\section{Numerical implementation}\label{appendix::numerical_implementation}
When fitting the proposed models to data we require accurate computation of the logarithm of the pdf $f_s(w)$. This requires some care and we detail some suggested methods.

\subsection{Uniform kernel}
For the uniform kernel, we require accurate computation of equation~\eqref{eqn::uniform_skewness}. We first rewrite this equation as
\begin{equation}
\begin{split}
    f_s(w) = \frac{b}{2a_+a_-}\Big\{&a_+\Big[\Phi\Big(\frac{bw'+a_-}{t}\Big)-\Phi\Big(\frac{bw'}{t}\Big)\Big]\\&+a_-\Big[\Phi\Big(\frac{bw'}{t}\Big)-\Phi\Big(\frac{bw'-a_+}{t}\Big)\Big]\Big\}.
\end{split}
\label{eqn::alternative_Uniform}
\end{equation}
Concentrating on the first part (as similar arguments apply to the second part) we require accurate computation of 
\begin{equation}
    \ln\Big[\Phi(x+c)-\Phi(x)\Big].
\end{equation}
We first compute 
\begin{equation}
    \ln\Phi(x)=-\ln2+\ln\mathrm{erfc}(-x/\sqrt{2}),
\end{equation} and note the following identity
\begin{equation}
    \ln\mathrm{erfc}(x)=
    \begin{cases}
    \ln\mathrm{erfcx}(x)-x^2&x>0,\\
    \ln\mathrm{erfc}(x)&x\leq0,
    \end{cases}
    \label{eqn::lnerfc}
\end{equation}
where $\mathrm{erfcx}(x)$ is the scaled complementary error function $\mathrm{erfcx}(x)\equiv\exp(x^2)\mathrm{erf}(x)$. An alternative computationally cheaper method to evaluate $\ln\Phi(x)$ uses the
\texttt{log\_ndtr} method implemented in \textsc{Scipy} which employs a Taylor series expansion for arguments $|x|>20$: typically this is a factor of two faster than evaluating $\ln\mathrm{erfc}(x)$.
% in equation~\eqref{eqn::uniform_kernel} 
We then compute the difference of two $\ln\Phi$ calls as
\begin{equation}
    \ln\exp(a-b) = a + \ln\Big(1-\exp(-(a-b))\Big),
\end{equation}
where we can evaluate the second term using a \texttt{log1mexp} function. For positive arguments we use the identity $\Phi(-x)=1-\Phi(x)$ and instead evaluate $-\Phi(-x)-\Phi(-(x+c))$.

For the second part of equation~\eqref{eqn::alternative_Uniform} we employ a similar procedure: for positive $(bw'-a_+)/t$ we evaluate $\Phi(-(bw'-a_+)/t)-\Phi(-bw'/t)$. The sum of the two parts can then be performed using a log-add-exp operation: $\ln\exp(a+b) = a + \ln\exp(b-a)$. 

\subsection{Laplace kernel}
For the Laplace kernel, we must evaluate equation~\eqref{eqn::laplace} which is unstable for large positive arguments. An alternative form using the scaled complementary error function $\mathrm{erfcx}(x)$
% \equiv\exp(x^2)\mathrm{erf}(x)$ 
is
\begin{equation}
\begin{split}
    f_s(w) = \sqrt{\frac{\pi}{8}}b\alpha\Big(\frac{bw'}{t}\Big)\Big[ \frac{1}{a_+}\mathrm{erfcx}&\Big(\frac{t^2-a_+bw'}{\sqrt{2}ta_+}\Big)
    \\&+\frac{1}{a_-}\mathrm{erfcx}\Big(\frac{t^2+a_-bw'}{\sqrt{2}ta_-}\Big)\Big].
\end{split}
\label{eqn::laplace_alt}
\end{equation}
This expression suffers from its own different numerical issues when the arguments are large and negative. Therefore, for optimum numerical stability we separate out the two terms in the sum and use the form in equation~\eqref{eqn::laplace} for large negative arguments and equation~\eqref{eqn::laplace_alt} for large positive arguments. Note that this occurs for $t^2-2abw'>0$ for one term and $t^2+2abw'>0$ for the other. We compute the logarithm of the two terms using equation~\eqref{eqn::lnerfc} and the similar identity
\begin{equation}
    \ln\mathrm{erfcx}(x)=
    \begin{cases}
    \ln\mathrm{erfc}(x)+x^2&x<0,\\
    \ln\mathrm{erfcx}(x)&x\geq0.
    \end{cases}
\end{equation}
The sum of the two terms can then be performed using a log-add-exp operation.

\section{Fornax models using the Gauss-Hermite series}
\begin{figure*}
    \centering
    \includegraphics[width=\textwidth]{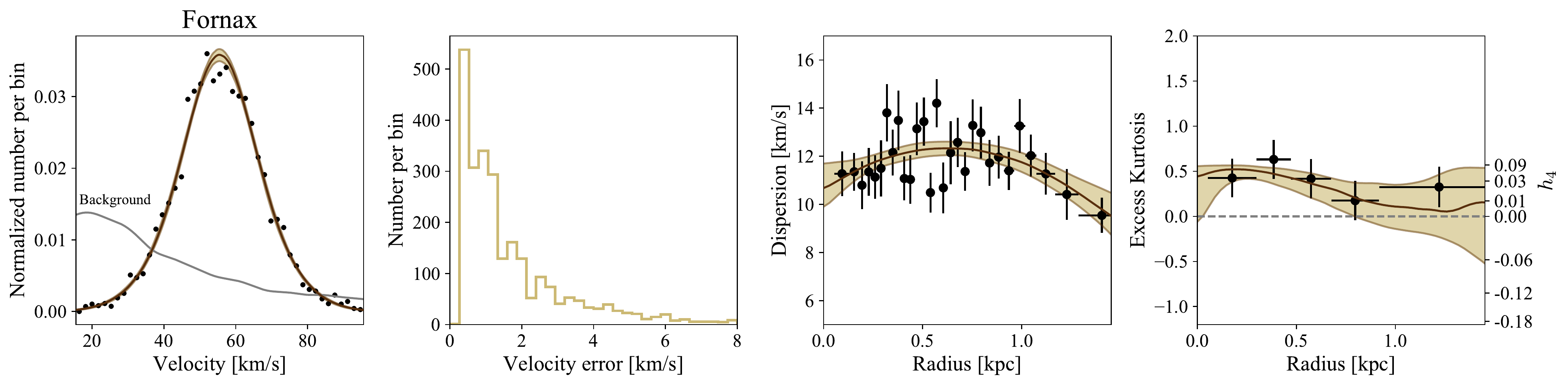}
    \caption{Similar to Fig.~\ref{fig:fornax_fit} but instead using a Gauss-Hermite series for positive excess kurtosis. Note in particular the different relationship between $h_4$ and the excess kurtosis for these models.}
    \label{fig:fornax_GH}
\end{figure*}
\begin{figure*}
    \centering
    \includegraphics[width=0.8\textwidth]{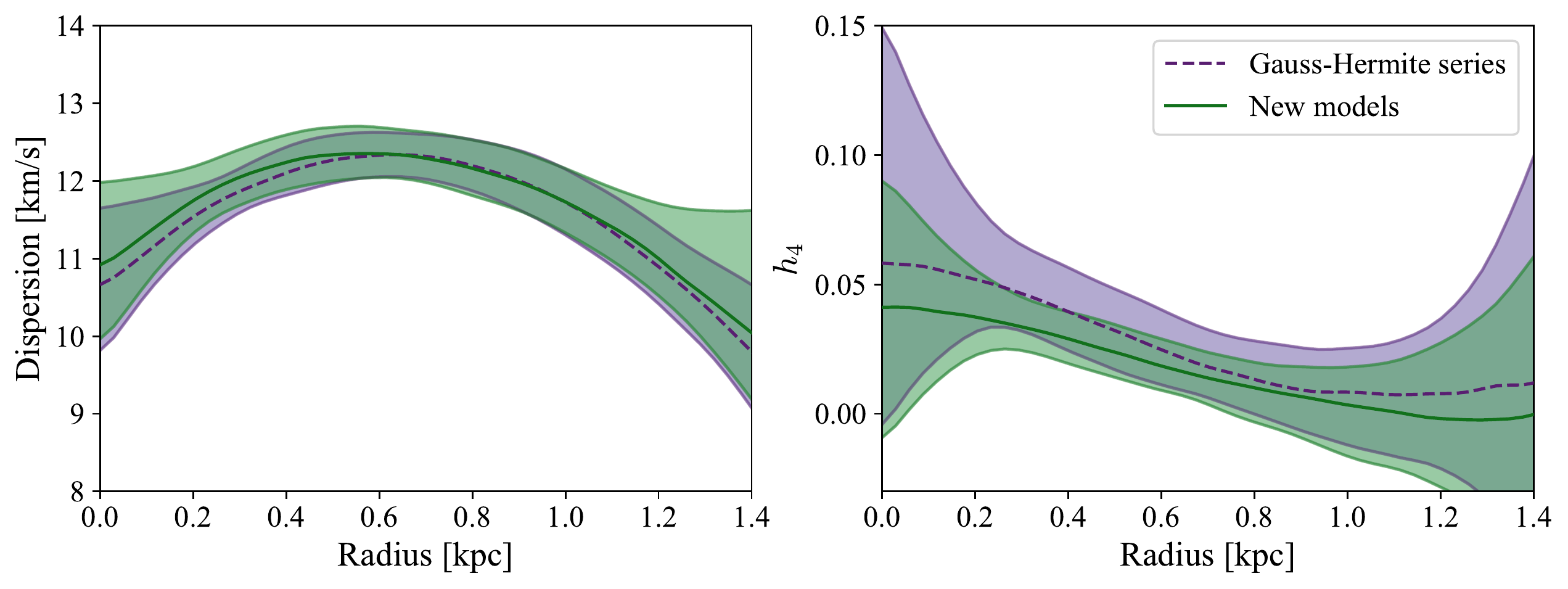}
    \caption{Comparison of the radial dispersion and $h_4$ profiles using the Gauss-Hermite series (dashed purple) and the Laplace kernel model (green solid) for positive $h_4$. The median and $\pm1\sigma$ bands are shown.}
    \label{fig:fornax_gh_comparison}
\end{figure*}
When working with models with positive excess kurtosis, it is possible to use the Gauss-Hermite series as it is always positive definite for $h_3=0$ and $h_4>0$. In Fig.~\ref{fig:fornax_GH}, we provide an alternative fit to the Fornax data using a Gauss-Hermite series instead of the Laplace kernel model introduced in this paper. Note that at equivalent $h_4$, the values of kurtosis admitted by the Gauss-Hermite series are smaller than for the Laplace kernel. However, as shown in Fig.~\ref{fig:fornax_gh_comparison}, both models produce similar dispersion and $h_4$ profiles.

% Don't change these lines
\bsp	% typesetting comment
\label{lastpage}
\end{document}